\documentclass[sigconf, screen]{acmart}
\setcopyright{none}

\usepackage{xcolor}  
\usepackage{listings}

\usepackage{multirow}
\usepackage{makecell}
\usepackage{cleveref}
\usepackage{float}
\crefname{figure}{Figure}{}
\crefname{section}{Section}{}
\usepackage{soul} 
\usepackage{graphicx}
\usepackage{subcaption}

\AtBeginDocument{%
  \providecommand\BibTeX{{%
    \normalfont B\kern-0.5em{\scshape i\kern-0.25em b}\kern-0.8em\TeX}}}

\acmConference[CHI '25]{CHI Conference on Human Factors in Computing 
Systems}{April 26-May 1, 2025}{Yokohama, Japan}
\acmDOI{10.1145/3706598.3713832}
\acmISBN{979-8-4007-1394-1/25/04}

\begin{document}

\title{PLAID: Supporting Computing Instructors to Identify Domain-Specific Programming Plans at Scale} 

\author{Yoshee Jain}
\authornote{Indicates that both authors contributed equally to the paper.}
\orcid{0009-0005-6291-5438}
\affiliation{%
  \institution{University of Illinois Urbana-Champaign}
  \city{Urbana}
  \state{Illinois}
  \country{USA}
}
\email{yosheej2@illinois.edu}

\author{Mehmet Arif Demirta\c{s}}
\authornotemark[1]
\orcid{0000-0001-5674-5878}
\affiliation{%
  \institution{University of Illinois Urbana-Champaign}
  \city{Urbana}
  \state{Illinois}
  \country{USA}
}
\email{mad16@illinois.edu}

\author{Kathryn Cunningham}
\orcid{0000-0002-9702-2796}
\affiliation{%
  \institution{University of Illinois Urbana-Champaign}
  \city{Urbana}
  \state{Illinois}
  \country{USA}
}
\email{katcun@illinois.edu}


\renewcommand{\shortauthors}{Yoshee Jain, Mehmet Arif Demirta\c{s}, \& Kathryn Cunningham}

\definecolor{editCol}{RGB}{0,0,0}
\newcommand{\edit}[1]{{\textcolor{editCol}{#1}}}

\begin{abstract}
Pedagogical approaches focusing on stereotypical code solutions, known as programming plans, can increase problem-solving ability and motivate diverse learners. However, plan-focused pedagogies are rarely used beyond introductory programming. Our formative study (N=10 educators) showed that identifying plans is a tedious process. To advance plan-focused pedagogies in application-focused domains, we created an LLM-powered pipeline that automates the effortful parts of educators' plan identification process by providing use-case-driven program examples and candidate plans. In design workshops (N=7 educators), we identified design goals to maximize instructors' efficiency in plan identification by optimizing interaction with this LLM-generated content. Our resulting tool, PLAID, enables instructors to access a corpus of relevant programs to inspire plan identification, compare code snippets to assist plan refinement, and facilitates them in structuring code snippets into plans. We evaluated PLAID in a within-subjects user study (N=12 educators) and found that PLAID led to lower cognitive demand and increased productivity compared to the state-of-the-art. Educators found PLAID beneficial for generating instructional material. Thus, our findings suggest that human-in-the-loop approaches hold promise for supporting plan-focused pedagogies at scale.

\end{abstract}



\begin{teaserfigure}
\centering
\Description{A three-stage figure showing how PLAID can be utilized. On the left, an instructor is looking at a set of LLM-generated programs and editing parts of the program to identify similarities. An arrow saying ``Identify programming parts in application focused-domains'' connects this to middle part, where multiple plans are shown as short code snippets with annotations. An arrow saying ``Support plan-focused pedagogies in that domain'' connects this to the right part, which shows different use cases for plans. These include `Organize course content', `Create examples', and `Build assessments'.}
\includegraphics[width=0.84\textwidth]{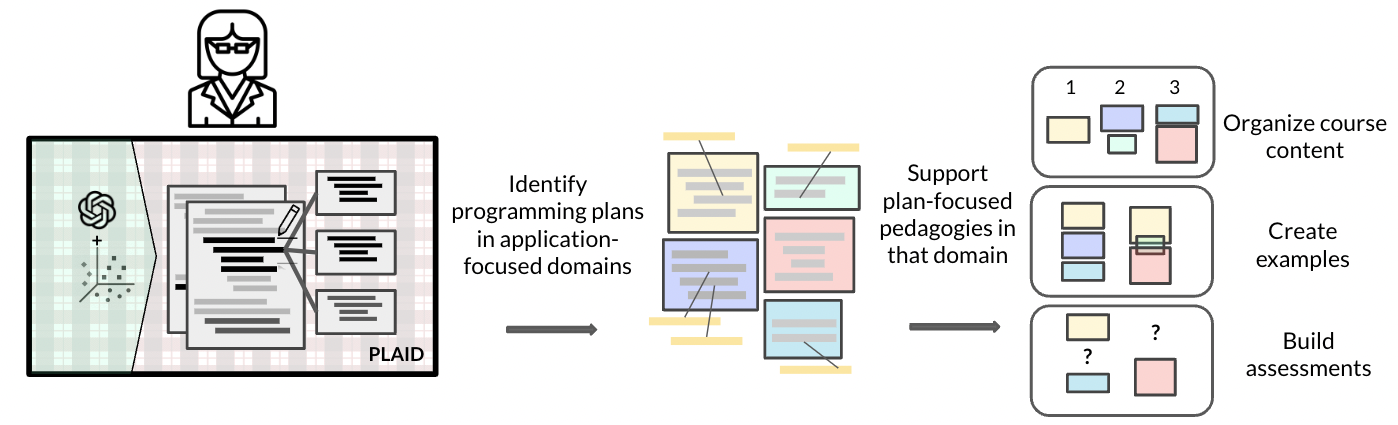}
\caption{PLAID supports instructors to more efficiently identify programming plans (i.e., common code patterns and information about their use) in application-focused programming domains by supporting their ability to explore and refine AI-generated content. Plan identification is a crucial step in the development of promising plan-based pedagogies. 
}
\label{fig:teaser-why-plaid-matters}
\end{teaserfigure}

\maketitle

\section{Introduction}









\begin{figure*}
\centering
\Description{Four example programming plans from prior work. The first one is captioned `Get a soup from multiple web pages', and shows a Python program with subgoal labels. Some parameters are replaced with fill-in-the-blanks boxes. The second one is captioned `Encapsulation in Object-oriented Programming', and shows a Python class definition with color coding for parts that could change. The third one is captioned `Find a value of the list that satisfies a certain condition', and a pseudocode solution is given with changeable parts highlighted. The fourth one is captioned `Arrange-Act-Assert Testing', and includes testing code with some parts highlighted.}
\includegraphics[width=\textwidth]{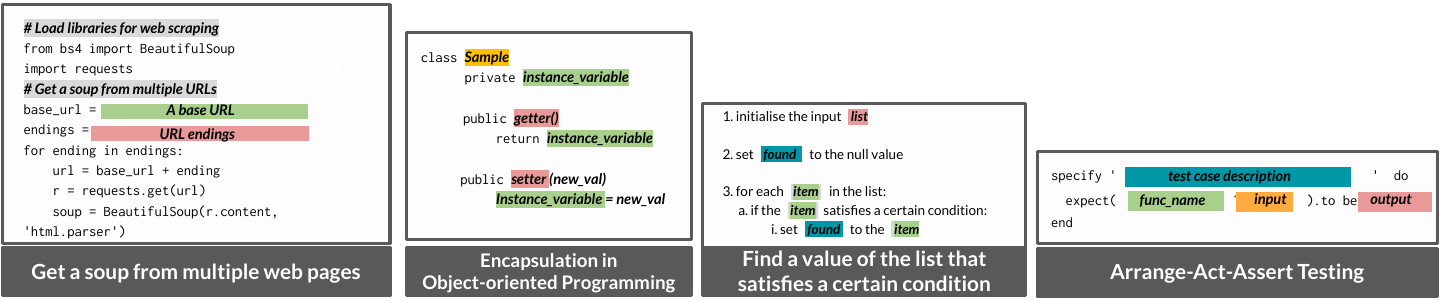}
\caption{Programming plans that educators have already identified from a variety of programming domains, including (from left to right) web scraping with Beautiful Soup~\cite{Cunningham_PurposeFirstProgramming_CHI-2021}, introductory object-oriented programming~\cite{Iyer_PatternCensus_SIGCSE-2021}, introductory procedural programming~\cite{Wermelinger_problems-to-programs}, and testing with RSpec~\cite{LojoFox_TestingPlan_ITiCSE-2022}.}
\label{fig:example-plans}
\end{figure*}

Programming plans, also known as programming patterns \cite{Iyer_PatternCensus_SIGCSE-2021, muller_pattern-oriented_2007} or code idioms \cite{haggis_code_similarity}, are short, stereotypical solutions that can be adapted to solve various problems with programming (see Figure~\ref{fig:example-plans}). Knowledge about programming plans has long been considered a significant aspect of programming expertise: 
the ability to recognize programming plans in code differentiates novices from experts~\cite{Soloway1984EmpiricalSO}, and knowledge of plans seems to underlie effective code writing~\cite{robins201912, spohrer_goalplan_1985}. 
When instructional strategies in computing education provide support in the form of programming plans, they can improve learners' ability to solve programming problems~\cite{Castro_ImpactSingleLecturePlans_Koli2017, muller_pattern-oriented_2007, Weinman_Fox_Faded_Parsons}. However, work that describes sets of programming plans~\cite{Spohrer1985-pi, Rist1989-fo, muller_pattern-oriented_2007, izu_inventory_2021, Iyer_PatternCensus_SIGCSE-2021} and that applies those plans in instructional tools~\cite{ Weinman_Fox_Faded_Parsons, Jigsaw} has so far been largely limited to only introductory programming content.

Recent research has shown that \textit{domain-specific} programming plans, which are plans from application-focused domains of programming such as data analysis or machine learning, can be effective in teaching and motivating a diverse set of students. Domain-specific plans have been used to support computer science majors learning software testing~\cite{LojoFox_TestingPlan_ITiCSE-2022}, expanding on the successful use of a similar tool in introductory programming~\cite{Weinman_Fox_Faded_Parsons}. 
A tool that guided students to learn and apply plans in a new topic area (web scraping with BeautifulSoup) was shown to motivate less confident ``conversational programmers''~\cite{Chilana_RiseConversationalProgrammer_VLHCC-2015} by giving them a brief and purpose-oriented overview of a new programming domain~\cite{Cunningham_PurposeFirstProgramming_CHI-2021}. 
These promising results imply that using programming plans in the instruction of a wider variety of programming topics can support learners in reaching their learning objectives.

\subsection{Challenges in Widespread Adoption of Plan-focused Pedagogies}
A major barrier in expanding plan-focused pedagogies to application-focused domains may be the difficulty of identifying the relevant programming plans in the first place. There are several challenges that prevent instructors from easily identifying these domain-specific programming plans.
One is the opaqueness of the plan identification process. Most computing education publications that identify plans 
do not describe their process and omit crucial details necessary to replicate the process in another domain. A more transparent process for plan identification may help experts identify plans in application-focused domains so they can better support learners interested in those applications. 
Secondly, what we do know about programming plan design suggests that instructors' state-of-the-art process is fully manual, and therefore potentially tedious and effortful. The cost of designing programming plans for more niche, application-focused domains may be deemed too high by instructors. A more efficient and supportive process could lower this barrier of entry, thereby enabling programming plan design for more instructors in diverse fields. 

There may be an opportunity for technologies to support domain-specific programming plan identification, but much is still unknown about how to best use these technological advances to support instructors in their plan identification process.
While there have been efforts in the software engineering community to automatically identify common code pieces~\cite{haggis_code_similarity, jezero_code_similarity, code_similarity_newest}, these approaches do not consider pedagogical aspects of identified constructs and likely do not meet the needs of instructors. However, with the latest developments in technologies such as large language models (LLMs) that can effectively explain and generate code~\cite{juryEvaluatingLLMgeneratedWorked2024a}, it might now be possible to build tools that support the identification of domain-specific programming plans by reducing the workload of instructors. 
However, it is currently unclear what the needs of instructors working on plan identification are, as well as how their process for designing programming plans can be improved.

\subsection{Introducing PLAID and its Design Process}
We present PLAID, a pedagogically-informed 
\textbf{\ul{PLA}}n \textbf{\ul{ID}}entification tool for instructors. 

To guide the design of PLAID based on the needs and challenges of instructors during the programming plan identification process, we conducted a formative study that included semi-structured interviews with ten computer science education researchers who had identified programming plans and used plans in instruction. We present insights into the current state-of-the-art in programming plan design, and we identify design opportunities for three major challenges instructors face in the design process.
Our findings suggest that simply identifying common code patterns is not sufficient for designing programming plans that meet pedagogical goals, preventing the process from being completely automated. 
Therefore, we propose a sociotechnical system that accelerates instructors' current design process
while standardizing the results
as the 
the best path to support plan-focused pedagogies in application-focused domains at scale. 


Through a series of design workshops with instructors, we contribute design goals for developing this sociotechnical system that assists instructors in identifying domain-specific programming plans.
Based on observations and feedback from seven instructors wit expertise in application-focused programming domains, 
we identify key interactions to support efficient viewing of reference material with many code examples, easy refinement of automatically generated candidate plans, support for exploring unfamiliar code functionality, and the ability to organize identified plans in a meaningful sequence.

We enact those design goals in PLAID, enabling instructors to view, understand, and edit a variety of LLM-generated domain-specific example programs and candidate plan content as they design programming plans in a standardized template. We show that the design of PLAID is effective based on a within-subjects evaluation study with 12 computing instructors that showed that instructors experienced lower cognitive demands and had a lower overall task load while using PLAID compared to a baseline condition representative of the current state-of-the-art. Instructors were able to design more plans and use their time more efficiently with PLAID, and they found using PLAID satisfying. Instructors liked having access to the large and diverse set of generated reference materials in the interface and were able to utilize those materials optimally to identify plans based on their own existing ideas as well as come up with novel plan ideas.
PLAID can be used to identify domain-specific plans in a wide variety of application-specific programming domains, supporting plan-based pedagogies at scale.

\subsection{Summary of Our Contributions}


This work explores how the capabilities of large language models can be utilized to support educators in the adoption of promising instructional approaches.
With PLAID\footnote{PLAID is available at http://tryplaid.web.illinois.edu/.}, LLM-generated reference content replaces the tedious parts of the plan identification process, allowing instructors to spend their effort primarily in areas that require pedagogical expertise: refining plans to better support the learning of their students. We show a concrete example of how LLMs can empower instructors by automating repetitive information-gathering tasks, allowing them to focus on tasks that require abstraction and other human expertise. 
This was achieved by treating instructors as key stakeholders throughout the design process, informing the design of all aspects of our tool with instructor needs, from the phrasing of the LLM prompts to the interactions with generated content.

Our contributions are three-fold: 
First, we gather insights into instructor practices for programming plan identification and the challenges they face
which has not been documented previously. Second, we identify opportunities to support plan identification and present design goals for incorporating LLM-generated content into instructor workflows for designing plan-focused instruction.
Finally, we design and evaluate a system that combines LLMs' strengths with instructor expertise, showing that it accelerates the design of programming plans and holds potential for increasing the adoption of plan-focused pedagogies in any given domain. 

\section{Related Work}


\subsection{Improving Instruction With Programming Plans}


Informed by schema theory, Elliot Soloway and his students first described \emph{programming plans} in the 1980s \cite{soloway1980problems,soloway-yale-studies,soloway1985problems}. They defined plans as chunks of code that achieve particular goals, like guarding against erroneous data or summing across a collection. 
They provided evidence that knowledge of plans can represent progress in learning programming~\cite{soloway1986learning}, and student errors could be explained in terms of misunderstanding plans \cite{looping,highfreq} and plan composition errors \cite{putalltogether}. 
Inspired by this, educational designers have used programming plans as scaffolding (assistance for learners~\cite{wood1976role}) to enable novice programmers to write more code than they would be able to on their own and designed plan-based curricula and tools~\cite{gpceditor, hohmann1992soda}. 
A variety of classroom instructional techniques have organized instruction around common patterns~\cite{clancy_designing_1996, muller_pattern-oriented_2007, muller_supporting_2008}, used plans to inform the design of instruction~\cite{Duran_ProgramComplexity_ICER-2018}, and evaluated students' expertise based on their ability to recall plans~\cite{arto_ava_reviewer}.

However, until recently, these instructional approaches were confined to introductory courses as most well-established sets of plans have been identified from introductory programming content (e.g., \cite{Iyer_PatternCensus_SIGCSE-2021, wallingford1998elementary, izu_inventory_2021}). Recent work proposed the purpose-first programming approach~\cite{Cunningham_PurposeFirstProgramming_CHI-2021} to support conversational programmers~\cite{Chilana_RiseConversationalProgrammer_VLHCC-2015, conversational-programmers-industry_Chilana_CHI-2016}, showing that plan-based scaffolding can support students with learning goals beyond those of an introductory programming course. While earlier systems focused on supporting introductory programming learning, purpose-first programming included plans drawn from an applied coding topic, web scraping, making programs from that application area accessible to novices. With a more efficient plan identification process, we can more easily extend promising pattern-based approaches like purpose-first programming to new programming application areas. 




\subsection{Identifying Patterns in Programs Automatically}

Software engineering literature on the mining of \textit{code idioms}~\cite{haggis_code_similarity, jezero_code_similarity, code_similarity_newest} uses statistical natural language processing to extract semantically meaningful pieces from a code corpus. This approach has been extended to support program synthesis using these idioms~\cite{shin2019idiomssynthesis}.
\edit{API development has been a fruitful context for the automated detection of programs with similar goals: static program slicing and machine learning techniques have been used to populate documentation of new APIs with relevant examples~\cite{montandon2013documenting, sohan2015spyrest} and statistical NLP has been used on OpenAPI data to suggest similar examples based on the content a programmer is currently developing~\cite{Moon_API-miner-API-to_API_FinanSE-2024}. } 
Techniques for representing code in ways that facilitate machine learning methods (e.g. code2vec \cite{alon2019code2vec}) expand the toolbox for code clustering methods. \edit{Some code summarization frameworks extract important lines of code as an intermediate step~\cite{sun2023extractiveandabstractiveframeworksourcecode}.}
\edit{While these approaches provide exciting methods for finding meaningful code patterns, all such techniques are intended for the software engineering community or experienced developers. Thus, they do not incorporate any pedagogical concerns that instructors may} consider when identifying programming plans, limiting their applicability to instruction. Understanding these pedagogical concerns 
\edit{may pave the way for} these natural language processing and machine learning approaches to be incorporated into instructional tools. 


Recent work in HCI has \edit{used some of the above methodologies to not only collect common patterns, but also display} commonalities in code samples in ways that support learning and instruction. Glassman et al. clustered student programming assignment data to reveal common learner misunderstandings in an interface built for instructors \cite{glassman2015studentsolutions}. 
Glassman et al. also visualized varieties of API calls from StackOverflow data~\cite{Zhang_StackOverflowAPIMining_ICSE-2018} to reveal common use cases to learners \cite{glassman2018visualizingapi}. \edit{CodeScoop~\cite{head2018interactive} eases the process of extracting complete and compilable code examples from larger programs to support developers sharing examples with their communities.} 
Although these \edit{systems may} help instructors, these works do not attempt to produce high-level code plans that emphasize the purpose of code over their implementation details. 
Forming concrete guidelines on how experts perform this abstraction process can inform the design of systems that support instruction.


\subsection{Incorporating Large Language Models Into Computing Education}

While large language models (LLMs) have not yet been used for identifying programming plans, their capability to generate code and related natural language descriptions has been widely established. In addition to solving programming exercises of varying topics and difficulty with high accuracy~\cite{finnie-ansleyRobotsAreComing2022, finnie-ansleyMyAIWants2023, wangExploringRoleAI2023}, LLMs have been employed to generate programming exercises in custom contexts~\cite{sarsaAutomaticGenerationProgramming2022b, logachevaEvaluatingContextuallyPersonalized2024} and novel assignment types that produce code using student explanations as input~\cite{dennyPromptProblemsNew2024, dennyExplainingCodePurpose2024}. Moreover, using LLMs to generate code explanations produced simpler and more accurate explanations compared to student submissions~\cite{leinonenComparingCodeExplanations2023}. Jury et al. further evaluated the code explanation capabilities of LLMs by generating worked examples, which are step-by-step solutions used for demonstrating the problem-solving process of an expert to a student, through expert assessment and a large-scale user study, showing that LLMs can not only generate code but also generate explanations on different levels of abstraction to explain the program to a novice learner~\cite{juryEvaluatingLLMgeneratedWorked2024a}. Despite these extensive capabilities, in the context of interactive design for education, the use of LLMs has been mostly limited to applications designed for \textit{students}, such as extensions for existing development environments~\cite{ferdowsiValidatingAIGeneratedCode2024, yanIvieLightweightAnchored2024, liTutorlyTurningProgramming2018}, programming assistants~\cite{kazemitabaarCodeAidEvaluatingClassroom2024, liffitonCodeHelpUsingLarge2024, yangDebuggingAITutor2024}, and interactive learning environments~\cite{jinTeachAIHow2024}.

Some studies also observed that LLM-generated code may include structures unfamiliar to novices~\cite{juryEvaluatingLLMgeneratedWorked2024a} or fail to follow industry best practices~\cite{ciprianoGPT3VsObject2023}. A potential research direction is designing tools to support \textit{instructors}, who might be better suited to address incorrect or inappropriate outputs generated by these models. 
\edit{While this area has not been explored extensively, studies have shown the promise of using LLMs for authoring learning objectives \cite{sridhar2023harnessing}, instructional content generation, grading, and question generation \cite{wang2024large}. \citet{mollick2023assigning}
explored the promise and pitfalls of AI as a tutor, coach, mentor, teammate, tool, simulator, and student, concluding that AI can be a ``force multiplier'' for teachers when used appropriately and in alignment with best practices. }
\edit{One such example of ``instructor-in-the-loop''} is VIVID~\cite{choiVIVIDHumanAICollaborative2024a}: a tool for helping instructors revise lecture videos into dialogue-style material. 
\edit{Another is TeachTune \cite{jin2024teachtunereviewingpedagogicalagents}, which incorporates a split-screen interface that allows instructors to evaluate and refine their pedagogical conversational agents. This interface enables users to build their agent using a graph-like state machine representation on one side while simultaneously observing the agent's real-time behavior with simulated students on the other.}
\edit{WEAT \cite{hassany2024humanaicocreationworkedexamples, hassany2024authoring} utilizes LLMs to generate first draft line-by-line explanations for code in worked examples which can then be edited by instructors. \citet{leiker2306prototyping} showed the potential of reducing human involvement in learning material generation for adult learners using LLMs.}

\edit{These tools} present promising precedent for systems that could help instructors generate content by presenting initial drafts of materials and supporting interactions for refinement. \edit{These tools can facilitate collaboration between the instructors and the LLM, leveraging LLMs' content generation capabilities and instructors' pedagogical expertise to create instructional content that follows best practices and aligns with learners' goals \cite{viola2023human}.} LLMs may be an appropriate candidate for automating parts of the plan identification process only if they are incorporated into a system that allows instructors to address their concerns with the generated content.



\begin{figure*}[h]
    \centering
    \Description{Three main parts of the study explained in the section, with research questions at top, methods at middle, and results at bottom levels.}
    \includegraphics[width=0.92\textwidth]{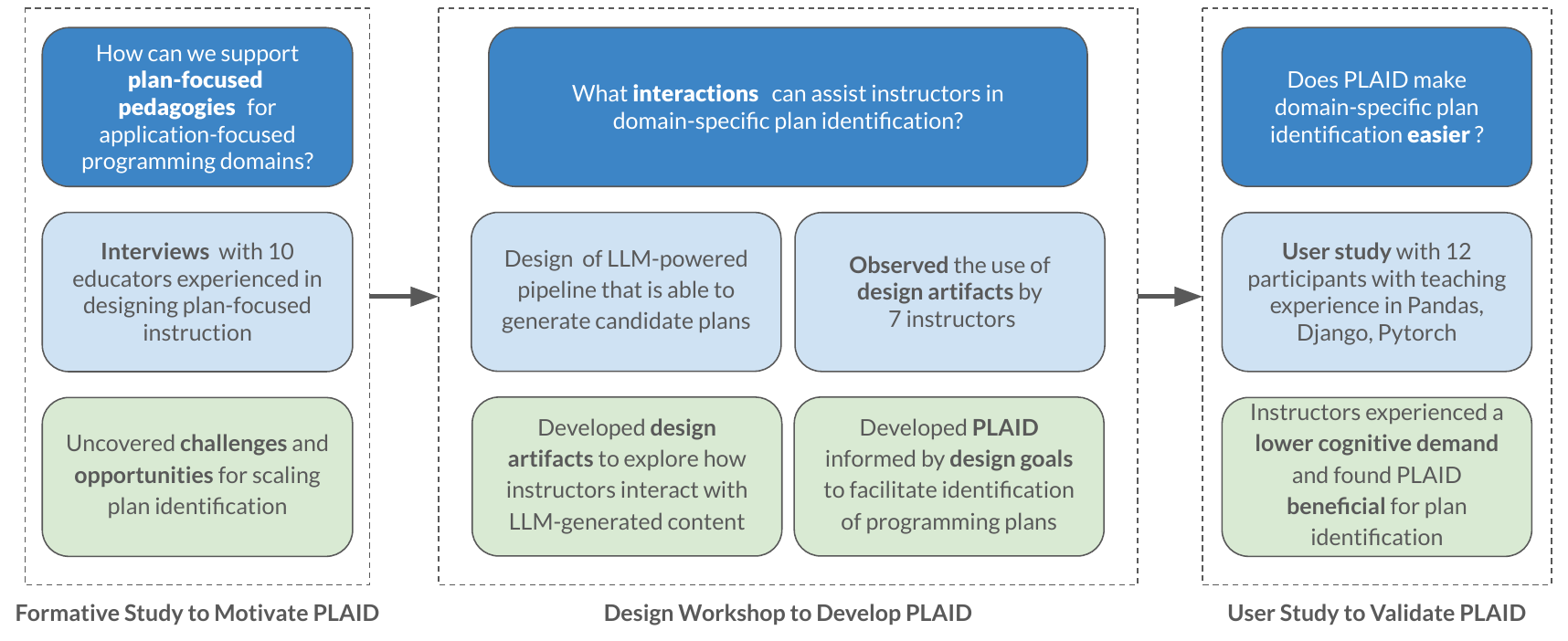}
    \caption{An overview of the three-step process to motivate, develop, and evaluate PLAID.}
    \label{fig:prelim-overview}
\end{figure*}

\section{Overview of Methodology}


We followed a three-stage design pipeline centering the needs of instructors throughout the  process
(see ~\cref{fig:prelim-overview}). First, to identify state-of-the-art techniques in plan identification and the key challenges educators face when using these strategies to design plan-focused pedagogies, we interviewed computing educators experienced in designing instructional material with programming plans.
We build a deep understanding of this previously undocumented process and uncover the key challenges for plan identification in \cref{sec:interview_results}.

Informed by these insights, we propose three key design characteristics that address these primary barriers. We evaluate these characteristics through a series of design workshops with instructors teaching application-specific domains of programming (e.g., data analysis with Pandas and web development with Django), validating their usefulness and revealing additional design considerations in \cref{sec:design-workshop}.

Enacting the identified design goals, 
we designed PLAID, a system that supports instructors in identifying plans in application-focused domains. We evaluated PLAID in a user study with twelve instructors and teaching assistants with varying teaching experience, finding that participants faced lower cognitive demands and were more productive compared to the baseline condition. We describe PLAID in \cref{sec:system-design} and report our user study in \cref{sec:user-study}.



\section{Challenges in the Process of Programming Plan Identification}
\label{sec:interview_results}
To understand the current process of programming plan identification by educators and its challenges, we conducted semi-structured interviews with ten computing instructors (see Table~\ref{tab:formative-participants}) who have identified lists of plans for use in instruction or understanding student skill development. 

\subsection{Study Design}
Mirroring the approach of Fowler et al.~\cite{craig_max_methoda}, we recruited instructors who authored a computing education research publication where plan identification was a part of their methodology or results. This allowed us to discuss their concrete programming plan identification experiences, rather than have participants speculate on the process in general. Computing education researchers are an ideal source to learn instructional best practices because this population consists largely of instructor-scholars: those who not only teach computing but are also informed by research and innovative practices in computing education.

We conducted online semi-structured interviews that ranged between 30 minutes and one hour. 
We asked participants to describe the specific activities they undertook to identify programming plans in the paper they authored. This included questions about what they were looking for when they identified programming plans, the resources that were involved in this process, and the procedures they undertook.
We also posed questions to gain insights into the difficulties the interviewees faced when they were identifying programming plans.
Lastly, we prompted a discussion about how they would attempt to identify programming plans in a new topic area they were unfamiliar with.



We used a transcription service to transcribe the video recordings and used Dedoose, a qualitative analysis software, for analyzing the transcripts with an inductive, reflexive coding process, influenced by thematic analysis approaches~\cite{clarke2021thematic}. During coding, we highlighted the connections between the beliefs of our participants and their choices in the pattern identification process. 
Note that while we use the term ``programming plan'' in this paper to clarify our focus on relatively small coding chunks rather than design patterns or architectural patterns, many of the educators we spoke with used ``plan'' and ``pattern'' interchangeably, or even preferred the term ``pattern.''

\begin{table}[h]
\caption{Demographics of the Instructor-Scholars Interviewed in our Formative Study.}
    \centering
    \footnotesize
    \label{tab:formative-participants}
    \begin{tabular}{l|cccccc}
    \toprule
            & \shortstack{Years \\ in CSEd \\ Research} & \shortstack{Years in \\ Plans \\ Research} & \shortstack{Years \\ in CS \\ Instruction} & \shortstack{Teaches \\ CS1?} & \shortstack{Uses \\ Plans in \\ Instruction?}
    \\\midrule
        P1 & 10-20 & 1-3 & 10-20 & Yes  & Yes \\
        P2 & 5-10 & 4-6 & 10-20 & Yes  & Yes \\ 
        P3 & 5-10 & 4-6 & 10-20 & Yes  & Yes \\ 
        P4 & 1-5 & 1-3 & 5-10 & No  & Yes \\ 
        P5 & 1-5 & 1-3 & 1-5 & Yes  & Yes \\ 
        P6 & 10-20 & 1-3 & 20+ & Yes  & Yes \\ 
        P7 & 5-10 & 4-6 & 20+ & No  & Yes \\ 
        P8 & 1-5 & 1-3 & 5-10 & Yes  & Yes \\ 
        P9 & 20+ & 20+ & 20+ & Yes  & Yes \\ 
        P10 & 20+ & 20+ & 20+ & Yes  & Yes \\ 
    \end{tabular}%
\end{table}

\subsection{Findings}
\label{sec:challenges}

Here, we describe the three most salient challenges relevant to the identification of domain-specific programming plans by instructors. These include difficulties these computing educators experienced while identifying programming plans for use in instruction, as well as a potential challenge for instructors new to the plan identification process. As we describe the challenges, we also include information about the plan identification process and approaches educators currently use to address these challenges.


\subsubsection{The Challenge of Finding Plans in the Practice (C1)} 
\label{sec:challenges_practice}
Our interviews confirmed that state-of-the-art programming plan identification is an entirely manual process. Echoing the belief that \textit{``patterns are `mined' from the practice''} (P10), a large part of that process included reviewing existing code, problem statements, or instructional material. This included GitHub repositories (P7), programs written by industry professionals (sometimes including their own code) (P7, P8, P10), textbooks (P1, P2, P5), and other instructional material including lecture notes, assignments, student programs, and testing material (P1, P2, P4, P6, P9).

While our participants had ready access to these resources, they still found it onerous to translate this practice into plans. As P7 described, \textit{``the challenge was trying to infer general characteristics from a large collection of specific examples.''} Another participant described this part of the process as \textit{``tedious''}, stating: 

\begin{quote}
``There was a lot of just paging through textbooks, either physical or digital versions and just looking at code, just trying to see if there's something we haven't seen before.'' (P1)
\end{quote}

Understandably, determining which aspects of a given example are widely applicable enough to ``count'' as a programming plan is a difficult task, as the possibilities for writing programs are so extensive. To manage this challenge, plan identifiers leaned on a combination of their own expertise (\textit{``you know it when you see it''} said P7) and collaboration with others.
This collaboration could involve discussions with co-instructors (P8, P9), TAs (P4, P8), study participants (P4), other researchers (P7, P9), and developers in the industry (P8). Recall that we interviewed authors of computing education publications, and it turns out that our interviewees frequently discussed plan identification with their co-authors (P1, P2, P3, P7, P9, P10) and even paper reviewers (P5).
These discussions occurred across different stages of the plan identification process, from when candidate plans are first presented, as well as when they are refined into their final form or removed from consideration. 

\subsubsection{The Challenge of Refining Plans for Student Needs (C2)} 
\label{sec:challenges_abstraction}
Once initial plans are chosen, the challenge only begins. Instructors judged the quality of the plans they identified based on multiple metrics, which could conflict. 
Not surprisingly, the instructors we interviewed generally agreed that a good programming plan should be used frequently in practice, whether that practice was classroom assignments (P1, P4, P8) or professional programming (P5, P8). A good plan \textit{``needs to be useful in many other contexts and ideally have some degree of flexibility to be adapted to similar situations"} said P3. 

At the same time, instructors believed that a good programming plan should be readily usable by learners, spanning the gulf between a problem statement and a code solution. P1 described plans as an \textit{``important step between understanding the syntax of a language and understanding how to do problem solving.''} P9 agreed, saying \textit{``knowing the pattern helps getting to a good solution, an expert solution.''} In the same way that interface designers narrow the ``gulf of execution'' between a user's goal and how to achieve that goal~\cite{Norman_UserCenteredSystemDesign_1986}, instructors believed that programming plans should facilitate learners' ability to design and implement coding solutions. \textit{``Understanding how to tune it should not be the issue,''} said P2. P10 agreed, stating \textit{``The solution has to be easy to put into practice (with practice).''}
 
Balancing the goals of a plan being \textbf{common in practice} while also being easily \textbf{usable by novice learners} was a major challenge for instructors who identified programming plans. 
P2 cautioned that \textit{``If it is too specific, it's not recurrent enough.''} On the other hand, P2 believed it is essential to ensure the plan is not \textit{``too abstract''} so that students \textit{``have difficulties in grasping the idea and using [the plans].''} Instructors described the challenge of balancing these factors as finding the right level of \textit{``abstraction''} (P2, P10), \textit{``generality''} (P2), or \textit{``granularity''} (P3). Ideally, instructors wanted to refine plans in a way that ensured they could be applied across multiple contexts, while also providing sufficient concrete detail to support learners to implement the plans correctly. 

Compounding this difficulty was the need to choose that level of abstraction appropriately for the learner audience. Most participants thought it is essential to \textit{``teach novices and experts patterns very differently''} (P1). This is because \textit{``advanced students can handle more abstract explanations''} (P10). 
To address the challenge of refining a plan to the correct level of abstraction, instructors employed an iterative process, involving discussions with colleagues (\textit{``A group of us would meet once or twice a year for several days and kick ideas about.''} said P10.). 

\subsubsection{Challenge of Ensuring Robust and Shareable Plan Definitions (C3)}
\label{sec:challenges_robust_shareable}
Our formative study revealed a fair bit of variety in the components of a programming plan that instructors looked for during the plan identification process. The instructors we interviewed mentioned one or more of the following components as crucial parts of a plan: (a) the \textit{name}, which is a short description of the plan (P1, P5, P6); (b) the \textit{goal}, which is a natural language phrase for what the plan achieves (P2, P3, P5, P6); (c) the \textit{solution}, which is how the plan is implemented, whether in code (P3), pseudocode (P1), or a description in natural language (P6); and (d) the \textit{changeable areas}, which show the places in the solution should be modified based on the specific context (P1, P6). These discrepancies present a challenge: plans identified by one instructor might not meet the needs or expectations of another. 

Our interviewees were highly knowledgeable about programming plans and plan-based pedagogies, but this is certainly not true of all computing instructors. Lack of knowledge about how to identify and use programming plans may be even more common among instructors who focus on application-specific programming areas and other topics beyond introductory programming, since the computing education research community as a whole tends to focus on introductory programming more than other programming topics~\cite{Noa_SIGCSE-focus-topics_2019}. Our sample of instructors was heavily biased towards CS1 instructors (see Figure~\ref{tab:formative-participants}), mirroring that trend. To meet the goal of supporting more instructors with more diverse content knowledge to design domain-specific programming plans, it may be necessary to constrain their experience and provide structure so they achieve similar outcomes to those of the plan identification experts we interviewed.

\section{Exploring Opportunities to Support Plan Identification in a Design Workshop}
\label{sec:design-workshop}



The challenges we identified in instructors' current plan identification process (Section~\ref{sec:challenges}) suggest that there are multiple opportunities to increase instructors' efficiency and ability to identify domain-specific plans. In this section, we describe arguments for three design characteristics that may improve the plan identification process.
Then, we report on the findings of design workshops in which instructors used and provided feedback about design artifacts with these characteristics. To ensure that our findings were applicable to domain-specific plan identification, we worked with instructors who teach application-focused programming domains and tailored their experience to include plan identification in those areas.
Finally, we report a set of design goals for systems that support domain-specific programming plan identification. 



\begin{figure*}[ht]
    \centering
    \Description{Illustration of three main characteristics of design artifacts. On left, a window annotated A shows a list of programs with short explanations on top. On top right, one of three clusters is shown to be selected, named `Summary Statistics'. On bottom right, an `Item Details' pane for the candidate plan obtained from cluster, annotated C, and the suggested potential values for plan components given on the right, annotated B.} 
    \includegraphics[width=0.8\linewidth]{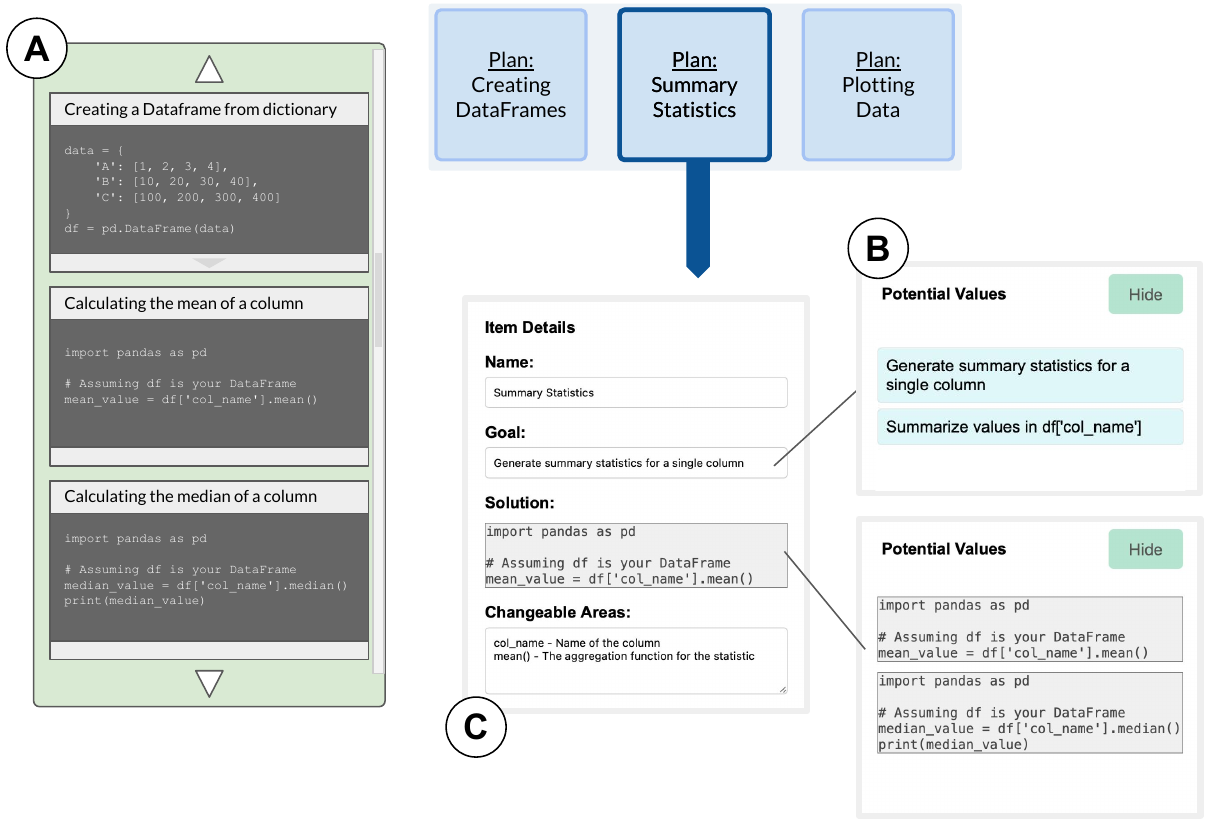}
    \caption{Illustrations of the three characteristics of the design artifacts, 
    proposed to support instructors' plan identification process by addressing the challenges from our formative study. 
    (A) Instructors can view a vast library of generated programs that achieve a diverse set of tasks in the relevant programming application area to inspire their plan identification; (B) Instructors can compare similar code snippets and other plan components to assist their refinement of plans; and (C) Instructors can follow suggested fields that endorse the structure their final plans should follow.}
    \label{fig:baseline-prototype}
\end{figure*}

\subsection{Characteristics of the Design Artifacts}
\label{sec:design-artifact}

The three key characteristics (see Figure~\ref{fig:baseline-prototype}) of our design artifacts are inspired by the three challenges identified in our formative study. These characteristics support a multi-stage workflow that mirrors educators' current plan identification process: initial identification of a plan candidates, refinement of the plan's details, and, finally, creating a complete description of the plan. 

\subsubsection*{(A) Support \textbf{Initial Plan Identification} with \textbf{Quick Exploration of Many Authentic Programs and Problems}} 
The participants in our formative study emphasized the importance of exploring content that captures a diverse set of authentic goals and implementations when identifying programming plans. However, they also found this search process to be tedious and time-consuming (Section \ref{sec:challenges_practice}). The instructors we spoke to indicated looking across multiple code files or examining textbooks to perform a survey of the topic in which they were identifying plans. \textit{If that reference material was readily available in a single interface, it may make the plan identification process quicker and easier.}
To address this challenge, we provide a large set of programs that address a diverse set of use cases in the target domain to reduce the tedium faced by instructors as they browse reference material and the burden of finding appropriate reference material for a certain programming domain.
These programs are generated by a large language model (LLM), described in full detail in Section~\ref{sec:llm-pipeline}. 

\subsubsection*{(B) Support \textbf{Plan Refinement} with \textbf{Comparisons of Similar Content}} 

Making decisions about exactly how to explain a particular concept in a programming plan was another common challenge highlighted by our formative study (see Section~\ref{sec:challenges_abstraction}). In refining their plan, instructors considered both how easily a potential plan could be used by their students (usability) and how common that plan was in practice (commonality). 
\textit{If instructors could view and compare multiple pieces of programs related to their potential plan, as well as a variety of potential plan goals, they may be able to more quickly evaluate how common or usable their plan is.}
We bootstrap this practice by clustering similar program snippets from the example programs, using a combination of heuristics and code embeddings (described in full detail in Section~\ref{sec:clustering}).

\subsubsection*{(C) Support \textbf{Robust and Shareable Plan Descriptions} with \textbf{Structured Fields}}
The final practice we focus from the plan identification process is refining initial candidates into the structure of programming plans. 
We found that, surprisingly, instructors did not agree on their preferred structure for a programming plan, presenting the potential challenge that plans identified by one instructor may not be usable by another (see Section~\ref{sec:challenges_robust_shareable}). We observed that plan identification requires iteration and refinement, and instructors have varied opinions on the components that they incorporate in their plans.
\textit{If instructors were given a structured template, they may iterate on their plans faster and achieve a convenient format for sharing with others.} By defining plans with all of the programming plan components mentioned by instructors in our formative study, they could ensure that plans they identify can be readily used by another instructor. Moreover, constraining instructors to work in such a structure could also support other educators who want to identify plans but are new to the pedagogical idea, potentially expediting their work by clarifying what exactly they should be looking for.
Thus, our interface enables examining plans with all their components for refinement rather than just code and comments in a general-purpose text editor or IDE.

\subsection{Workshop Methodology}

To understand whether the design characteristics described accelerate domain-specific programming plan identification, as well as to identify what interactions with a plan identification system are valued by instructors, we hosted a series of design workshops.
We invited seven instructors with expertise in a variety of application-focused programming domains to participate (see Table~\ref{tab:participants-design-workshop}).
These areas included data analysis with Pandas (four instructors), web development with Django (two instructors), and web scraping with BeautifulSoup (one instructor). We did not require prior plan identification experience from participants.



\subsubsection{Protocol.}
Each instructor completed a screen- and audio-recorded 90 minute session involving three plan identification tasks and an interview. Participants were compensated with a \$75 Amazon gift card. 
Before the tasks, participants received a brief overview of the definition of programming plans and their components, based on the findings of our formative study (see Section~\ref{sec:challenges_robust_shareable}). 
We told participants that they should create plans appropriate for students with some Python familiarity but no experience in the application-focused domain (i.e., Pandas, Django, or Beautifulsoup).
Then, participants proceeded to perform plan identification in their domain of expertise across three conditions.
In each condition, participants were asked to create up to five programming plans during a 15 minute window in a thinkaloud setting. 


Our goal was to observe (a) how instructors interacted with the three characteristics in a low-constraint environment akin to paper prototyping~\cite{Sefelin_PaperPrototyping_CHIEA-2003} (Conditions 1 and 2), and (b) how instructors interacted with our more highly-constrained prototype (Condition 3). Across these conditions, we hoped that instructors' actions and feedback might suggest useful interactions to best make use of our proposed characteristics, as well as information about whether our prototype supported plan identification as expected. 
\begin{itemize}
    \item Condition 1: a Miro board\footnote{https://miro.com} populated with 30 example programs from the domain and associated one-line explanations in natural language. In addition, 70 more example programs could be found in a linked Google Sheet. Instructors were asked to create their plans in boxes labeled with each of the plan components. \textit{(Characteristics A and C)},
    \item Condition 2: a Miro board populated with common code snippets from the example programs, clustered into groups by similarity and ranked by frequency. Suggested goals, names, and changeable areas were also listed with the code snippets. Instructors were asked to create their plans in boxes labeled with each of the plan components. \textit{(Characteristics B and C)}, 
    \item Condition 3: a web application that supports navigation through the example programs from Condition 1, viewing of "suggested plans" that contained clustered code snippets and associated goals, names, and changeable areas from Condition 2, and a plan creation area where plans can be edited within the suggested fields. \textit{(Characteristics A, B, and C)}
\end{itemize}
Participants were informed that they were not required to use provided content and could access external sources at any time, including web searches, library documentation, their own code, or their teaching material.
The sessions ended with a short interview, during which the participants were asked to rank all three conditions based on how instructor-friendly they were, make requests for potential new features, and describe their likelihood of using a similar interface to generate programming plans for their courses. 

\begin{table}
\caption{Demographics of the Instructor Participants in our Design Workshop.}
    \centering
    \footnotesize
    \label{tab:participants-design-workshop}
    \begin{tabular}{l|ccccc}
    \toprule
            & Domain & \shortstack{Teaching \\ Experience \\ in CS} & \shortstack{Teaching \\ Experience in \\ Domain} & \shortstack{Used \\ Plans in \\ Instruction?}
    \\\midrule
        W1 & Pandas & 16-20 & 1-5 & No \\
        W2 & Django & 6-10 & 1-5 & No \\
        W3 & Pandas & 6-10 & 1-5 & Yes \\
        W4 & BeautifulSoup & 6-10 & <1 & Yes \\
        W5 & Pandas & 1-5 & 1-5 & No \\
        W6 & Django & 16-20 & 1-5 & No \\
        W7 & Pandas & 6-10 & <1 & Yes \\
    \end{tabular}%
\end{table}


\subsection{Findings}
\label{sec:design-workshop-findings}

\subsubsection{Condition 1: Interactions with Example Programs from the Application-Specific Domain.}
Most participants reviewed the example programs and use cases before designing any plans. Interviewees indicated that they found the provided examples to be meaningful and authentic (``\textit{this idea of merging datasets is really really valuable}'', said W1). Participants found value in both the code and the associated natural language descriptions. W7 added that reading ``\textit{titles are probably more useful than the code}.'' Instructors used these examples as inspiration for ideas about the different concepts that they can build plans around.

W2 and W7 explicitly communicated that designing one plan inspired them with ideas for new plans, rather than sticking to a list of ideas they came up with at the beginning of the session. 

However, participants did not always find what they were looking for. W1 and W4 had particular ideas for designing their plans, but they could not find the implementation of those ideas in the reference materials, leading to time wasted in a fruitless search. While most participants were positive about the provided examples, some preferred to gather inspiration from their own teaching experience or an external resource. Specifically, W5 preferred to design plans using their own prior work (a course website with modules for Pandas), and W7 consulted the Pandas documentation for clarifications and ideas. 

In this condition, participants created plans with different structures from one another.
Most notably, participants annotated changeable areas in different ways: highlighting parts of code using a different color or changing the text color (W1, W6), drawing rectangles or ellipses around parts of changeable code (W4, W7), or writing a natural language description of changeable parts (W1, W2, W3, W4, W5).

Nonetheless, the process of plan identification was not straightforward. Sometimes, in the given reference material, participants saw syntax that they were unfamiliar with. In this case, instructors performed web searches to clarify their understanding of the structures. For instance, W4 redirected to StackOverflow to find the difference between \textit{``\texttt{.text} and \texttt{.content}''} and clarified that \textit{``there's ..., not a substantive difference''}. W5 asked questions about the ``\textit{specificity}'' of plans, and W6 requested clarifications about the intended audience. In addition, participants found it challenging to understand the difference between the plan names and goals. W2 articulated that names might be less meaningful to students, even if instructors do get it.
Despite the hurdles, instructors reacted positively to 
this condition, praising the existence of reference materials. They ranked this condition second highest among the three.

\subsubsection{Condition 2: Interactions with Clusters of Code Snippets and Suggested Plan Components.}
\label{sec:workshop-findings-condition2}
In contrast to condition A, participants often struggled to create plans with the reference material provided in condition B and found the experience ``\textit{overwhelming}'' (W1) or tedious because they had to navigate a large canvas (W4).
More importantly, participants expressed dislike of the content itself, which consisted of clusters of similar small code snippets as well as suggested goals, names, and changeable areas. Many participants preferred to include complete worked examples rather than code snippets in their final plans, with the idea that it would help students to avoid any need for \textit{``implied knowledge ''} (W1). Similarly, some instructors were concerned about the size of the snippets, finding them ``\textit{too fine-grained''} (W4). 
W4 felt the suggested names of plans were ``\textit{too generic}'' and that many of the code snippets presented the same ideas repeatedly.
While in a few cases, instructors were able to find inspiration among the clusters of code snippets, this  condition received the lowest ranking by all but one participant.

\subsubsection{Condition 3: Prototype with Improved Navigation of Content.}

Participants found the
interactions in the prototype to be useful, and this condition was ranked the highest by most interviewees.  
While identical reference information as in earlier conditions was presented, the features of the prototype made it easier for instructors to navigate and use the given content. W4 indicated that the interactions for browsing candidate plans reduced the effort in reviewing many clusters in previous conditions. W2 and W7 appreciated the two-column view of example programs, where they could scan the list of natural language descriptions in one column  and click to review code on the other column if they find that use case interesting. W3 felt that the features of the prototype improved their efficiency over the earlier conditions:
\begin{quote}
    The Miro board itself doesn't really bring any value....
    I better understand the ways where things like the use case palette and the library [in the prototype] could be used to quickly do the compositing. 
\end{quote}

At the same time, participants requested ways to further reduce their search space, such as searching for important keywords (W2).
While the prototype addressed some concerns of instructors in the other conditions, our interviewees did request that some features of Miro boards be incorporated (e.g., highlighting code to emphasize changeable areas (W1)). When asked about additional features, almost all interviewees requested syntax highlighting on code examples and snippets, similar to the format of code editors.

W2, a Django instructor, found it challenging to represent plans in a single template as many Django programs are spread over multiple files. In such domains, visualizing connections between multiple plans could be helpful.


\subsection{Design Goals}
\label{sec:design-goals}

Reflecting on our findings from this workshop, we formulate four design goals to inform the development of PLAID (or other systems) that assist instructors in designing programming plans with LLM-generated content.

Mirroring the process of reviewing code from Section~\ref{sec:challenges_practice}, we observed that instructors valued the availability of many complete example programs for inspiration and brainstorming.
However, the ways in which instructors interacted with these examples varied based on their personal values. 
Some instructors emphasized the importance of having contextualized examples in their plans to motivate students. 
Others focused on incorporating best practices or conventional techniques that 
are important to remember. 
So, PLAID should be able to present instructors with \textbf{authentic} code examples that capture context and key functionality at the same time.
Some instructors also described multiple approaches to instantiate the same plan, emphasizing the different ways of achieving the same goal. 
PLAID should provide \textbf{diverse} examples to instructors to help them capture these different approaches.
In addition to having access to a corpus of programs, the content should be \textbf{displayed in a compact view} to enable instructors to easily navigate and search for key concepts.

\begin{quote}
    \textbf{DG1: PLAID should inspire instructors by presenting many diverse and authentic examples in a compact view.}
\end{quote}

While instructors have expertise in their target domains, the dynamic nature of domain-specific libraries (e.g. Pandas) makes it hard for experts to keep up with up-to-date conventions and practices.
In our design workshop, when participants saw strange syntax, they used external resources like library documentation and web searches to understand unfamiliar code. 
So, PLAID should incorporate mechanisms that can help instructors \textbf{understand} unfamiliar programming paradigms. 
Moreover, most computing instructors write programs using code editors where they can frequently run code to review their implementations. So, PLAID should allow instructors to \textbf{view the output} of their programs to validate and refine their material.

\begin{quote}
    \textbf{DG2: PLAID should assist instructors in mitigating uncertainties with example code.}
\end{quote}

\begin{figure*}[h]
        \Description{Three annotated screenshots, showing the buttons and menus explained in the text.}
        \includegraphics[width=\textwidth]{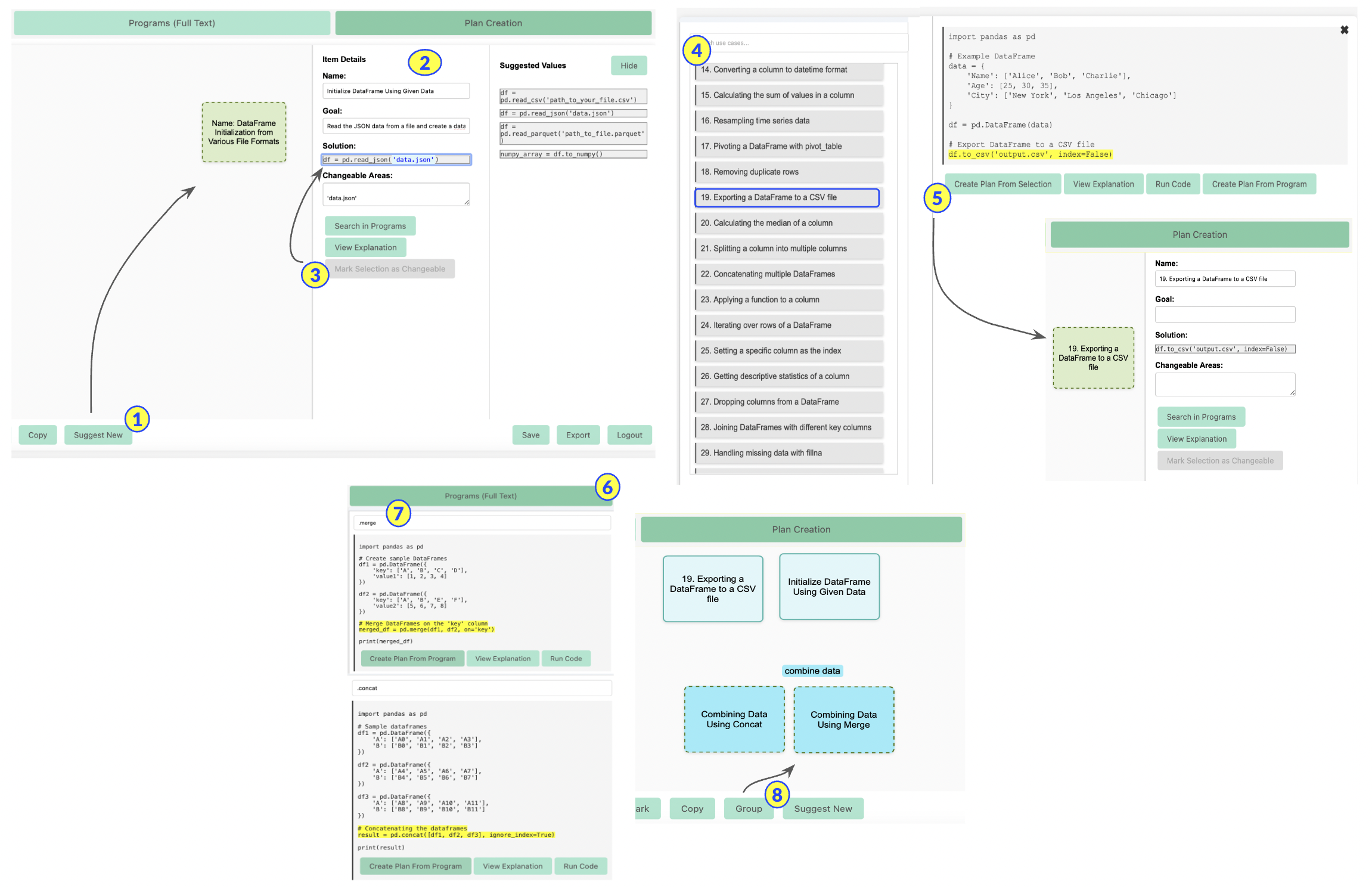}
        \caption{Jane's Workflow Diagram. Jane (1) asks the system to suggest a plan; (2) edits the name and goal components of her plan, (3) marks the changeable areas in the plan, (4) browses the list of use case to find code that meets her goals, (5) selects the relevant part of the code from the full program and creates a plan from the selection, (6) switches to the full programs tab to search for specific pieces of code, (7) uses the search bar to search for keywords, and (8) groups plans with similar goals.}
        \label{fig:jane-workflow}
\end{figure*}

A key challenge indicated by instructors was the tiresome nature of combining reference content from many different sources to design plans. Exploring multiple sources was distracting, and navigating various platforms reduced their efficiency. Moreover, we saw that some instructors preferred to use the given material as an initial draft and refine it rather than write code from scratch. 
So, PLAID should support efficient interactions for editing reference material to \textbf{speed up} plan refinement. Participants indicated that it is challenging to abstract high-level ideas from multiple similar programs, so PLAID should \textbf{support abstraction} by allowing instructors to quickly combine content from a variety of sources.

\begin{quote}

    \textbf{DG3: PLAID should accelerate instructor workflows with efficient interactions for abstracting concepts and designing plans.}


\end{quote}

To adopt plan-based pedagogies in courses, instructors need to make decisions about how to organize and present plans to students. 
In our design workshop,
we observed that participants attempted to document how they would use plans during instruction by clustering plan boxes on the canvas by theme. They also ordered plans in the sequence that students should learn about them.
So, PLAID should present instructors with ways of organizing content, as they \textbf{navigate their pedagogical concerns} about how students should be learning about plans.

\begin{quote}
    \textbf{DG4: PLAID should support instructors in organizing content to address pedagogical concerns.
    } 
\end{quote}










\begin{figure*}
    \centering
    \Description{PLAID's system architecture diagram. Top part shows the database (a), and bottom part shows the interface (b). The system starts from bottom right as an instructor is interested in a programming domain, then the pipeline described in the text produces reference materials at different levels of granularity, and these are presented in the interface.}
    \includegraphics[width=\textwidth]{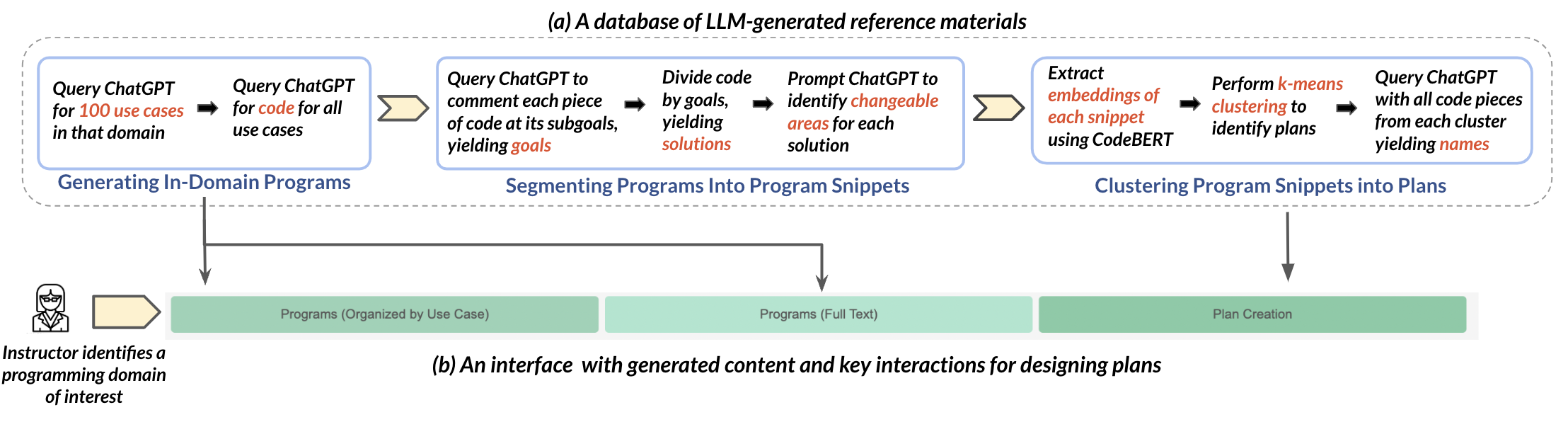}
    \caption{PLAID's reference content is generated through an LLM pipeline
    that produces output on three levels.
    First, a wide variety of use cases are generated to create example programs that focus on code's applications. Next, using LLM's explanatory comments that represent subgoals within the code, the examples are segmented into meaningful code snippets. The LLM is queried to generate other plan components for each code snippet. Finally, the code snippets are clustered to identify the most common patterns, representing plan candidates. The full programs are presented in `Programs' views of PLAID interface, whereas snippets are presented in clusters in the `Plan Creation' view.}
    \label{fig:system-pipeline}
\end{figure*}
\section{PLAID: A System for Supporting Plan Identification}
\label{sec:system-design}

Following the design goals devised from the design workshop, we refined our early prototype into PLAID: a
tool to assist instructors in their plan identification process.
PLAID synthesizes the capabilities of LLMs in code generation with interactions enabling plan identification practices observed in our studies with instructors.
As we noted in the findings of our design workshop, the LLM-generated candidate plans are not ready to be used as is in instruction, but instructors can readily adapt and correct them (\cref{sec:workshop-findings-condition2}).
PLAID enables collaboration between instructors and LLMs, enhancing the plan identification process by automating its time-intensive information-gathering tasks and facilitating instructors' ability to refine LLM-generated candidate plans based on their knowledge about pedagogy and the programming domain.

\subsection{Practical Illustration}

To understand how instructors use can PLAID to more easily adopt plan-based pedagogies, we follow Jane, a computer science instructor using PLAID to design programming plans for her course (summarized in \cref{fig:jane-workflow}).

Jane is teaching a programming course for psychology majors and wants to introduce her students to data analysis using Pandas. As her students have limited prior programming experience and use programming for specific goals, she organizes her lecture material around programming plans to emphasize purpose over syntax. 

She logs in to the PLAID web interface, 
and asks PLAID to suggest a plan (\cref{fig:jane-workflow}, 1). The first plan recommended to her 
is about reading CSV files. 
She thinks the topic is important and the solution code aligns with her experience; 
but she finds the generated name and goal to be too generic. She edits (\cref{fig:jane-workflow}, 2) these fields to provide more context that she feels is right for her students.
To make this plan more abstract and appropriate for more use cases, 
she marks the file path as a changeable area (\cref{fig:jane-workflow}, 3), generalizing the plan for reading data from different file formats.

Inspired by the first plan, she decides to create a plan for saving data to disk. She wants to teach the most conventional way of saving data, so she switches to the use case tab (\cref{fig:jane-workflow}, 4) and explores example programs that save data to get a sense of common practices.  
She finds a complete example where a DataFrame is created and and saved to a file. 
She selects the part of the code that exports data to a file and creates a plan from that selection (\cref{fig:jane-workflow}, 5).

For the next plan, she reflects on her own experience with Pandas. She recalls that merging DataFrames was a key concept, but cannot remember the full syntax. 
She switches to the full programs tab (\cref{fig:jane-workflow}, 6) that includes complete code examples and searches (\cref{fig:jane-workflow}, 7) for ``\texttt{.merge}'' to find different instances of merging operations. 
After finding a comprehensive example, she selects the relevant section of the code and creates a plan from it.

After designing a set of plans that capture the important topics, she organizes them into groups (\cref{fig:jane-workflow}, 8) 
to emphasize sets of plans with similar goals but different implementations. For instance, she takes her plans about \texttt{.merge} and \texttt{.concat} and groups them together to form a category of plans that students can reference when they want to {combine data from different sources}.


She exports these plans to support her students with minimal programming experience by preparing lecture slides that organize the sections around plan goals, using plan solutions as worked examples in class, and providing students with cheat sheets that include relevant plans for their laboratory sessions.


\begin{figure*}[h]
        \Description{An annotated screenshot of PLAID's `Programs' view. On the left, a list of use cases such as `Renaming columns in a Frame' and `Plotting a histogram of a column' is shown, with a scrollable list and a search bar. The latter one is selected, and on the right, we see the contents of the program in a monospaced font, with four buttons explained in the caption.}
        \includegraphics[width=\textwidth]{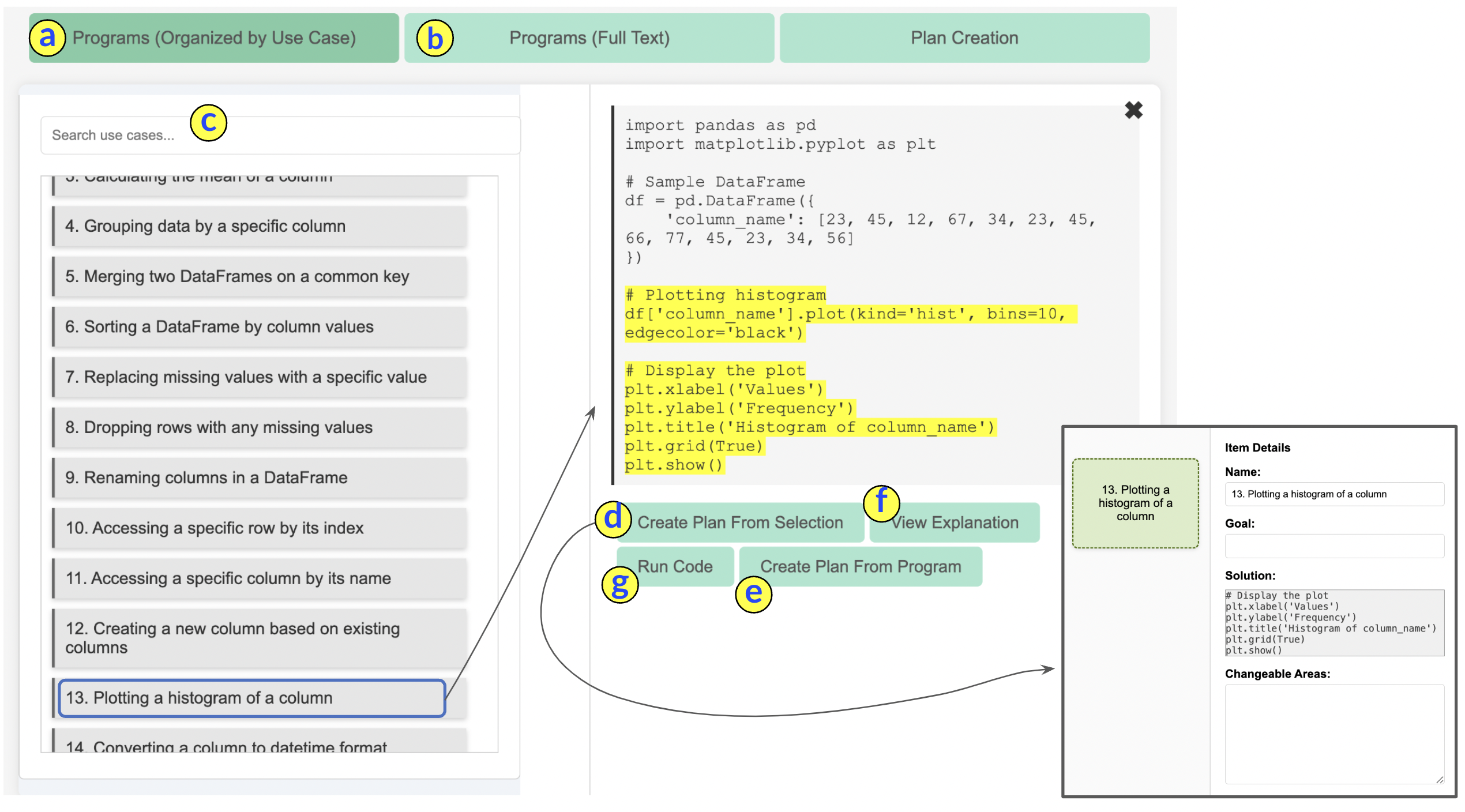}
        \caption{Plan Identification using PLAID: (a) list of example programs for instructors organized by natural language descriptions, (b) list of full programs of code, (c) search bar enabling easy navigation of given content to find code for specific ideas, (d) button to create a plan using the selected part of the code, (e) button to create a plan using the complete example program, (f) button to view an explanation for a selected code snippet, and (g) button for executing the selected code.}
        \label{fig:system-diagram-1}
\end{figure*}

\subsection{System Design}

\begin{figure*}[h]
    \Description{An annotated screenshot of PLAID's Plan Creation view with three panes, with plans shown as boxes on the left. A plan is highlighted, and we see its components on the middle pane. On the rightmost pane, we see suggested values for the selected component.}
    \includegraphics[width=\textwidth]{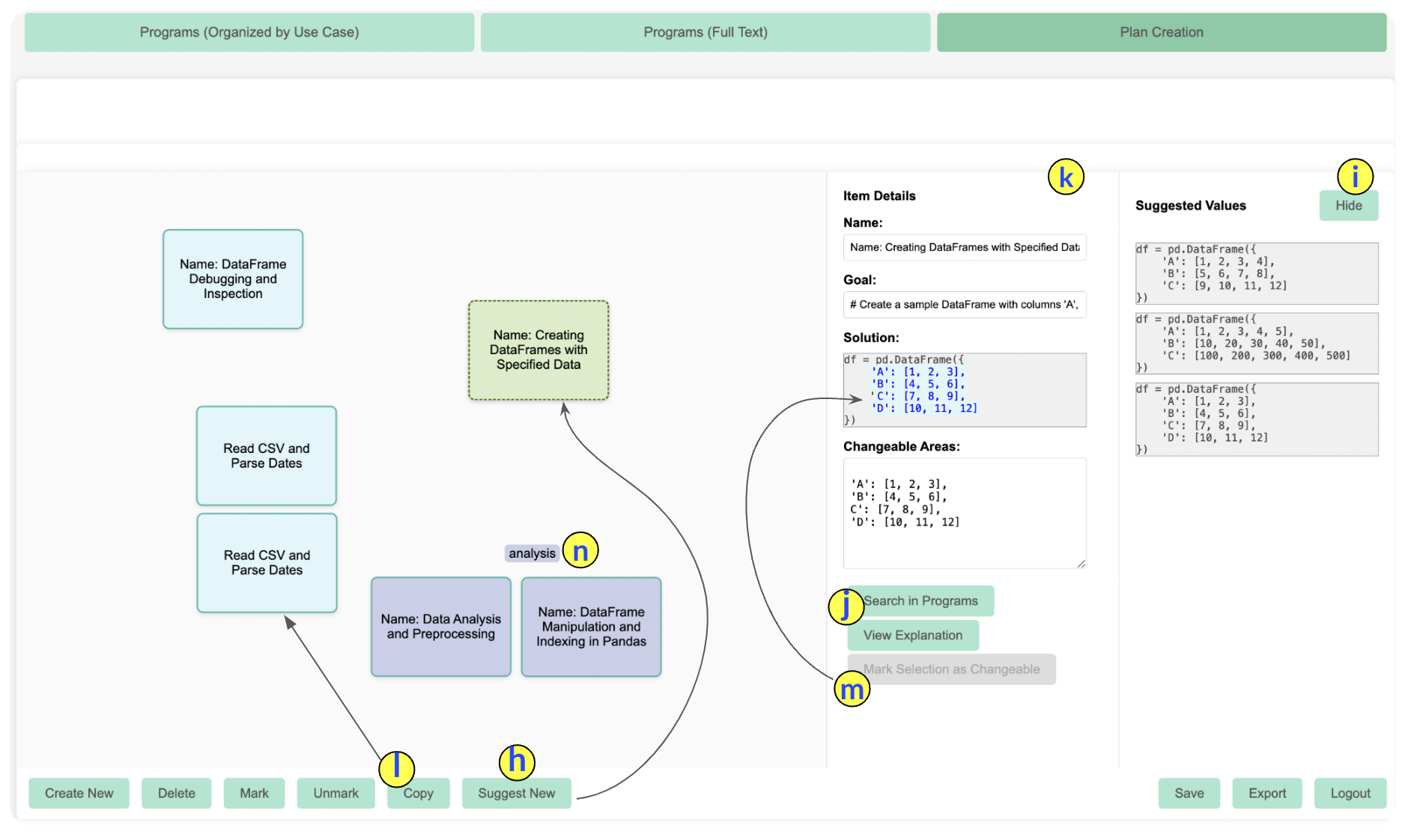}        
    \caption{Plan Identification using PLAID: (h) button that suggests a domain-specific candidate plan from the system database, (i) pane enabling viewing of similar values for the selected plan component, (j) button to view the solution code as part of a complete program, (k) pane with a structured template for plan design with editable fields to refine plan components, (l) button to copy a selected plan, (m) button to mark snippets of code from the plan solution as changeable areas, and (n) a button to group plans together into a category and add a name.}
    \label{fig:system-diagram-2}
\end{figure*}

At a high level, PLAID\footnote{The code for PLAID can be found at: https://github.com/yosheejain/plaid-interface.} operates on two subsystems: (1) a database of LLM-generated reference materials created through a pipeline that uses \edit{OpenAI's GPT-4o\footnote{https://openai.com/index/hello-gpt-4o/}~\cite{achiam2023gpt}}, inspired by instructors' best practices for identifying programming plans (see ~\cref{fig:system-pipeline})
and (2) an interface that allows instructors to browse reference materials for relevant code snippets 
and refine suggested content into programming plans
(see Figures~\ref{fig:system-diagram-1} and~\ref{fig:system-diagram-2}).

\subsubsection{Database of Reference Materials for Application-Focused Domains}

PLAID extracts information from reference materials at three levels of granularity to support each instructor's unique workflow: complete programs that address a particular use case, annotated program snippets that include goals and changeable areas, and plan candidates that cluster relevant program snippets together.

\textbf{Generating complete example programs.}
The content at the lowest level of granularity in the PLAID database are the complete programs. 
\label{sec:llm-pipeline}
As these examples should capture a variance of use cases in the real world, we utilized an LLM trained on a large corpus of computer programs and natural language descriptions~\cite{liu2023isyourcode}.
We prompted\footnote{Full prompts can be found in \cref{sec:appendix-pipeline}.} the model to generate \texttt{specific use cases of <application-focused library>}, defining use case as \texttt{a task you can achieve 
with the given library} (see \cref{sec:use_case_prompt}). Subsequently, we prompted the model to \texttt{write code to do the following: <use case>}, producing a set of 100 example programs with associated tasks (see \cref{sec:code_prompt}). By generating the use cases first and generating the solution later, we avoided the problems with context windows of LLMs where the earlier input might get `forgotten', resulting in the model producing the same output repeatedly. For practical purposes, we generated 100 programs per domain. \edit{To test for potential ``hallucinations'' where the LLM generates plausible yet incorrect code~\cite{Ji_2023_hallucination}, we checked the syntactic validity of the generated programs before developing the rest of our pipeline. No more than one out of 100 generated programs included syntax errors in each of our domains, i.e., Pandas, Django, and PyTorch. Thus, we concluded that hallucinations are not a major threat to the code generation aspect of PLAID.}

\textbf{Generating annotated program snippets.}
The second level of granularity in PLAID consists of small program snippets and a goal, with changeable areas annotated. 
We used the generated programs from
the prior step as the input to the LLM to add subgoal labels, where we prompted the LLM to annotate subgoals (see \cref{sec:subgoals_prompt}) as comments that describe \texttt{small chunks of code that achieve a task that can be explained in natural language}. These subgoal labels were used to break the full program into shorter snippets. Each snippet was fed back to the model to generate changeable areas (see \cref{sec:ca_prompt}), defined in the prompt as \texttt{parts of the idiom that would change when it is used in different scenarios}. The subgoal label that explained a code snippet corresponded to its goal in the plan view and the list of elements assigned as changeable was used for annotations.
\textbf{Generating clustered plan candidates.}
\label{sec:clustering}
The highest level of granularity provided in PLAID
are
plan candidates, in the form of clusters of annotated program snippets. To compare the similarity of program snippets, we used CodeBERT embeddings following prior work~\cite{codebert} and applied Principal Component Analysis (PCA) \cite{PCAanalysis} to reduce the dimensionality of the embedding while preserving 90\% of the variance. The snippets were clustered using the K-means algorithm~\cite{kmeansclustering}, using the mean silhouette coefficient for determining optimal K~\cite{silhouettecoeff}. Each cluster is treated as a plan candidate, with the goal, code, and changeable areas from each program snippet in the cluster presented as a suggested value for the plan attributes.
For each plan candidate, a name (see \cref{sec:name_prompt}) that summarizes all snippets in the cluster was generated by prompting an LLM with the contents of the snippets and stating that it should generate \texttt{a name that reflects the code's purpose} and it should focus on \texttt{what the code is achieving and not the context}. 





\subsubsection{Interface for Designing Programming Plans}
Building on the 
design goals identified in the design workshop (\cref{sec:design-goals}), PLAID enables a set of key interactions to assist instructors in refining candidates to design plans for their instruction.

\textbf{Interactions for Initial Plan Identification.}
While instructors valued the availability of code examples in the design workshop (Section~\ref{sec:design-workshop-findings}), we observed many opportunities for scaffolding their interaction with the reference material. To this end, PLAID presents example programs in two different views \textbf{(DG1)}. 
The ``Programs (Organized by Use Case)'' (\cref{fig:system-diagram-1}a) tab includes a list of use cases where instructors can click on an item to expand the program for that use case.
The ``Programs (Full Text)''  tab (\cref{fig:system-diagram-1}b) lists all the programs and enables instructors to scroll or search through (\cref{fig:system-diagram-1}c) all the code at once.
Both views support directly creating a plan from the whole example (\cref{fig:system-diagram-1}e), or a selected part of it (\cref{fig:system-diagram-1}d), by copying the solution and the goal of the program into an empty plan template
further supporting efficient use of the material \textbf{(DG3)}.

To facilitate understanding unfamiliar code and syntax, we implemented a ``View Explanation'' button (\textbf{DG2}) that generates a short description of the selected line(s) of code by prompting an LLM (\cref{fig:system-diagram-1}f). 
Participants also looked for code execution to validate and understand a program. However, since the code snippets instructors work with are often incomplete in this task, we implemented a ``Run Code'' feature (\textbf{DG2}) that predicts the output of a selected code snippet by prompting an LLM to walk through the code \texttt{step by step}, using Chain-of-Thought prompting~\cite{wei2022chain} (\cref{fig:system-diagram-1}g). Only the predicted output for the code is presented, ignoring other output from the LLM.


\textbf{Interactions for Plan Refinement.}
To provide suggestions for code patterns common enough to be potential programming plans,
we utilize the clustered program snippets from our database. In the ``Plan Creation'' view of PLAID, instructors can ask for suggestions (\cref{fig:system-diagram-2}h) to see a candidate plan to refine (\textbf{DG3}).  \edit{This functionality allows instructors to draw on their experience to recognize common code snippets and decide if they are valuable to teach students.}
\edit{This promotes recognition over recall \cite{recognition_over_recall}, thus helping reduce the cognitive effort that instructors may have to put in while designing programming plans traditionally.}
To allow instructors to better understand the context of a plan under refinement, PLAID 
also includes a button for searching for the current solution within the entire set of full programs
(\textbf{DG3}, \cref{fig:system-diagram-2}j).

As instructors edit the components of a plan, they are shown similar values from the corresponding component in that cluster (\cref{fig:system-diagram-2}i). By clicking on any suggested value, instructors can replace a plan component with a suggestion that better captures that aspect of the plan \textbf{(DG1)}. \edit{By allowing instructors to view the plan they are working on along with other related code pieces in a split screen view, we promote instructor efficiency by reducing the split-attention effect \cite{tarmizi1988guidance}. In the current plan creation process, even when using LLMs from their chat interface, instructors would have to switch between windows with code examples and their text editor which may increase the load on the instructors' working memory \cite{clark2023learning}. In PLAID, instructors can edit their plans and view similar code pieces at the same time.}



\textbf{Interactions for Building Robust and Shareable Plan Descriptions.}
PLAID encourages instructors to design plans in a structured template (\cref{fig:system-diagram-2}k). Moreover, PLAID reinforces the plan template by providing a dedicated method for annotating changeable areas by highlighting any part of the code (\textbf{DG3}, \cref{fig:system-diagram-2}m). Instructors can further explain the changeable areas by adding comments as text.

Our design workshop showed that participants would create a plan and copy it to emphasize alternatives or modifications to the underlying idea. To support this workflow,
PLAID allows users to ``duplicate'' plans on the canvas and further edit them to present alternative solutions for the same plan \textbf{(DG3}, \cref{fig:system-diagram-2}l).

To encourage instructors to think about organizing plans in ways that they would present them to students, PLAID provides an open canvas view for instructors that allows them to arrange plans as they prefer. In addition, PLAID supports a ``grouping'' feature (\cref{fig:system-diagram-2}n), which allows instructors to combine plans with similar goals together into one category (\textbf{DG4}).


\subsubsection{System Architecture}
The pipeline to create reference materials is implemented in Python, using the state-of-the-art large language model GPT-4o (Model Version: 2024-05-13). The interface for PLAID is implemented as a web application in Python as a Flask webserver, with an SQLite database. The user-facing interface is implemented using HTML, CSS, and JavaScript, with the canvas interactions realized with the library `\textit{interact.js}'.

\section{Evaluation of PLAID}
\label{sec:user-study}

To evaluate PLAID, we aimed to determine if computing instructors were able to use PLAID to identify plans in an application-specific programming domain more efficiently and with a more positive user experience than the current state-of-the-art. 
Specifically, we performed a within-subjects user study to gather insight into (1) instructors' productivity in the plan identification process, (2) the task load for using the system, and (3) the overall usability of PLAID. 

\subsection{Study Design}
\subsubsection{Participants}
\edit{The target end-users for PLAID are computing instructors who intend to create instructional content to teach an application-specific computing course. So, we} recruited four instructors and eight graduate teaching assistants with at least a year of experience in teaching \edit{a programming course (see Table~\ref{tab:participants-evaluation}) at the undergraduate level whether in introductory or upper-level programming courses}. In addition to teaching experience, our inclusion criteria required participants to indicate expertise in at least one application-focused programming domain: data analysis with Pandas (six participants), machine learning using Pytorch (four participants), and web programming with Django (two participants). None of our participants had prior experience identifying plans for instruction. Each participant engaged in a 60-minute design session and was compensated with a \$50 Amazon gift card.

\subsubsection{Procedure}
We conducted a within-subjects study, where each instructor performed plan identification with a baseline condition representing the current state-of-the-art and with PLAID. The study was counterbalanced, with half of our participants seeing the baseline condition first and the other half seeing PLAID first.

Each session began with a description of what a programming plan is, using an example from introductory programming. Then, participants were given 15 minutes in each condition to identify programming plans in their application-focused domain. We prompted them to create these plans as if they will be used in lectures that teach important concepts to students with no experience in their domain. To encourage instructors to undertake a significant amount of plan identification, they were given a suggested target of four to five plans. They were encouraged to continue if they reached this goal before their time was over. 
After each condition, participants completed the NASA-TLX questionnaire~\cite{cao2009nasa} to indicate their workload while performing the task. 

\textit{Baseline condition.} Participants worked on an empty Google document. They were given one example programming plan from introductory programming, which they could refer to to understand the expected plan structure. Participants were allowed access to external resources, including web searches, ChatGPT or other AI tools, or their own code and content.

\textit{PLAID condition.} Participants were given access to the PLAID web interface after a short demonstration of fundamental interactions supported by the system by the interviewer. Like the baseline condition, they were allowed to access any external resources besides the content suggested by PLAID.

\subsubsection{Post-task reflection.}
 
The sessions ended with a short interview, asking participants for feedback on the system and their opinions on plan-based pedagogy.
Participants also evaluated PLAID with the PSSUQ Version 3 usability survey \cite{pssuq_usability, sauro2016quantifying}. 

\begin{table}
\caption{Demographics of the Participants in our User Study.}
    \centering
    \footnotesize
    \label{tab:participants-evaluation}
    \begin{tabular}{l|cccccc}
    \toprule
            & Domain & Academic Title & \shortstack{Teaching \\ Experience \\ in CS} & \shortstack{Teaching \\ Experience \\ in Domain} & \shortstack{Used \\ Plans in \\ Instruction?}
    \\\midrule
        E1 & Django & Instructor & 20+ & 1-5 & No \\
        E2 & Pandas & Instructor & 1-5 & 1-5 & No \\
        E3 & Django & Instructor & 11-15 & 6-10 & No \\
        E4 & Pytorch & Graduate TA & <1 & <1 & No \\
        E5 & Pandas & Instructor & 1-5 & 1-5 & No \\
        E6 & Pytorch & Graduate TA & 1-5 & <1 & No \\
        E7 & Pandas & Graduate TA & 1-5 & <1 & No \\
        E8 & Pytorch & Graduate TA & 1-5 & <1 & No \\
        E9 & Pandas & Graduate TA & 1-5 & <1 & No \\
        E10 & Pytorch & Graduate TA & 1-5 & <1 & No \\
        E11 & Pandas & Graduate TA & 1-5 & 1-5 & No \\
        E12 & Pandas &  Instructor & 20+ & 6-10 & No \\
    \end{tabular}%
\end{table}

\begin{figure}[h]
    \centering
    \Description{A box plot with seven pairs of horizontal bars. Each bar corresponds to one of the measures on NASA TLX, with the top bar being the overall score. Median value for PLAID is better than baseline for all measures, and the difference is significant for overall score, physical demand, and mental demand.}
    \includegraphics[width=\linewidth]{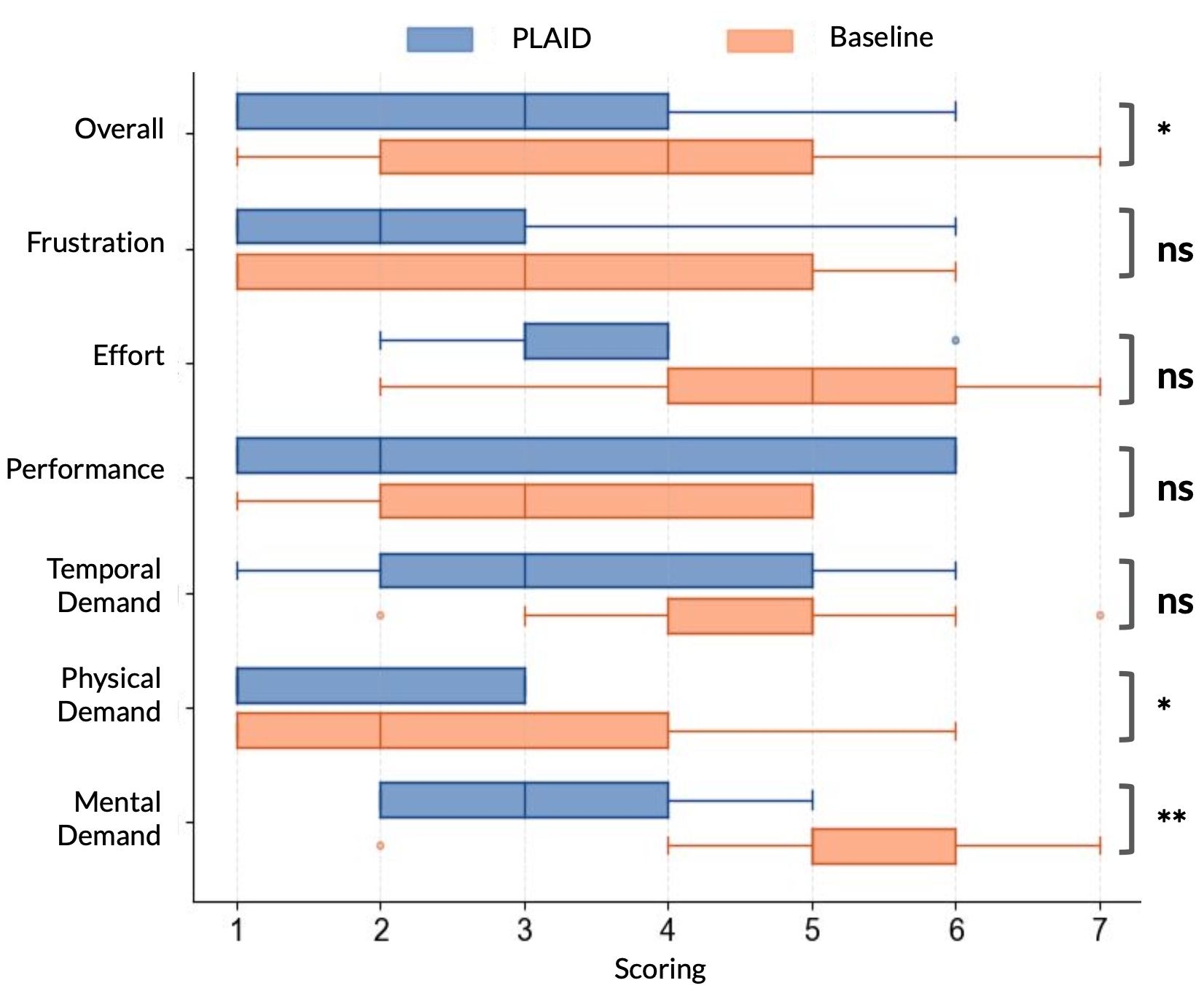}
    \caption{Participants' responses on the NASA Task Load Index survey administered after both the baseline condition and PLAID condition. For all items, lower values are preferred. The chart also indicates the results of the Wilcoxon signed rank test between the baseline and PLAID conditions. $**$, $*$, $ns$ indicate $p < 0.01$, $p < 0.05$, and $p > 0.5$ respectively.}
    \label{fig:cognitive-load}
\end{figure}

\begin{figure*}[h]
    \Description{Bar plots showing the distribution of responses for each item across system usability (SYSUSE), information quality (INFOQUAL), and interface quality (INTERQUAL) on the PSSUQ survey. For most items, The majority of items are rated above then the median option (Neither Agree or Disagree).}
    \includegraphics[width=\textwidth]{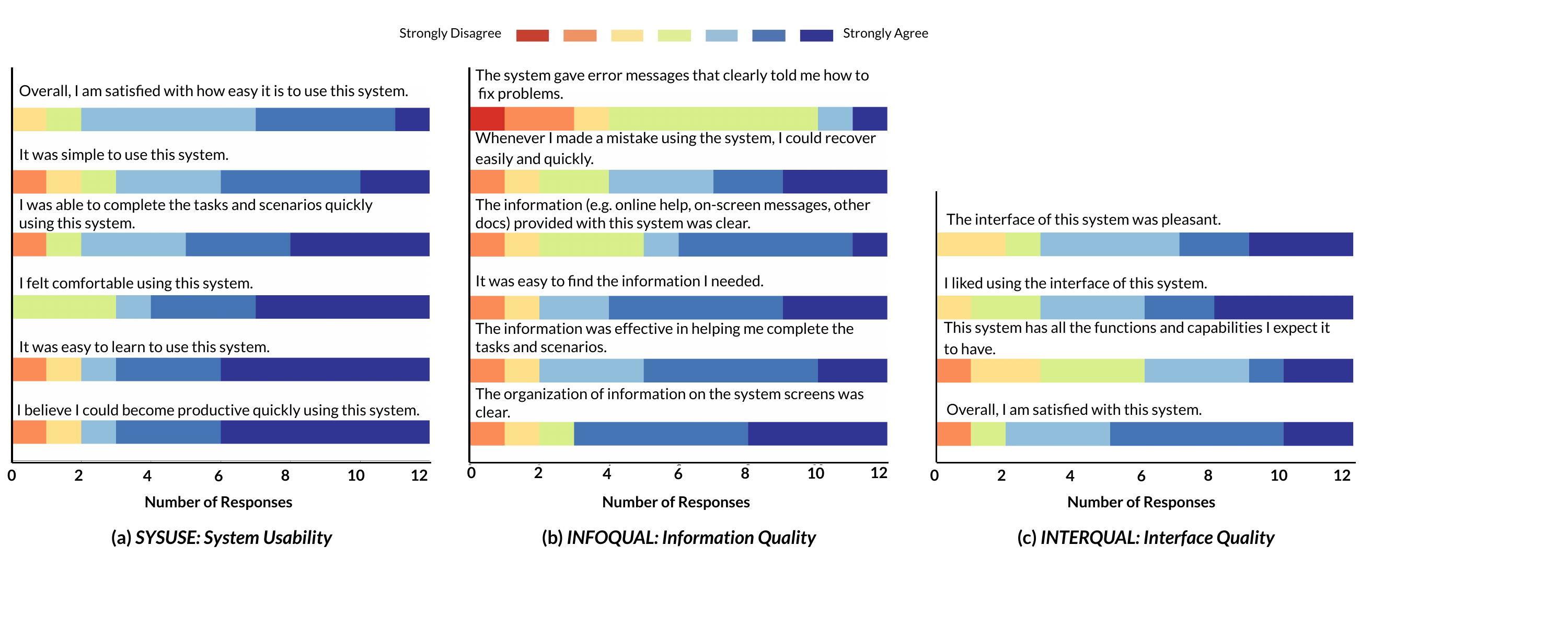}
    
    \caption{Self-reported reflections of participants on the usability of PLAID using the PSSUQ survey. The graph encapsulates their responses for each question across each category on a 7-point Likert Scale.}
    \label{fig:three-horizontal}
\end{figure*}

\subsection{Findings}


\subsubsection{PLAID enables instructors to identify plans more efficiently.}
Participants created more plans when using PLAID (4.75 plans on average) compared to the baseline condition (3.92 plans on average).
Seven participants were able to reach the target of identifying five plans in the PLAID condition, whereas only three of twelve participants were able to identify five plans in the baseline condition.


However, there was a wide variety in participants' ability to identify plans, which could be impacted by many factors, such as the individual instructor's content knowledge, the particular domain they are working in, the condition they are in, the condition they started with, and how much time they spent doing plan identification so far.

To understand how other experimental factors affect the time instructors take to identify plans, we built a linear mixed-effects model with the time spent per plan as the outcome variable. Fixed effects were the experimental condition (baseline or PLAID), session order (started with baseline or PLAID), how far into the task the instructor is (number of plans they have identified before this plan), and their domain (Pandas, Django, or PyTorch). The participants were modeled as random effects to control for differences in their expertise and other individual values. Even with a small sample size of 12 instructors and 104 identified plans, we observed a marginally significant coefficient for the experimental condition ($b = -57.2 (sec), t=-1.77, p = .079$) when controlling for these other factors, indicating that instructors were faster in identifying plans using PLAID by almost one minute per plan compared to the baseline. We also observed a statistically significant difference between the specific plans within a task and the time taken to design each plan ($b=129.2 (sec), t=12.85, p < .001$), indicating that instructors spent more time designing each plan for the later plans they suggested, potentially due to starting with easier concepts and moving to more complex ones.




\subsubsection{PLAID decreases cognitive demands and overall task load during plan identification.}

We find that the average task load for instructors\footnote{Computed across E2 to E12. E1 was excluded due to procedural error.} was significantly lower with PLAID (\cref{fig:cognitive-load}), indicated by Wilcoxon signed-rank test ~\cite{wilcoxon1992individual} (PLAID: $M = 2.83$ , $SD = 1.40$, Baseline: $M = 3.94$, $SD = 1.57$, $p = .04$). In addition, differences in two sub-measures were statistically significant: mental demand (PLAID: $M = 3.09$, $SD = 1.04$, Baseline: $M = 5.18$, $SD = 1.32$, $p = .008$), and physical demand (PLAID: $M = 1.63$, $SD = 0.92$, Baseline: $M = 2.54$, $SD = 1.75$, $p = .047$).

\subsubsection{PLAID provides instructors with a satisfactory experience.}
Participants responded positively to the PLAID user experience as indicated by responses to the PSSUQ survey\footnote{Option 1 indicated strong agreement and Option 7 indicated strong disagreement.} items (see Figure~\ref{fig:three-horizontal}). The responses aggregated into an overall mean of $M = 2.73$ ($SD= 1.49$), $M = 2.42$ for System Usefulness (SYSUSE, $SD= 1.46$), $M = 2.99$ for Information Quality (INFOQUAL, $SD = 1.57$), and $M = 2.81$ for Interface Quality (INTERQUAL, $SD = 1.44$).



\begin{figure*}[t!]
    \centering
    \Description{A visualization of participant actions in the system. Each participant is represented as a horizontal sequence of square markers, and the color of the marker corresponds to one of these actions: Create Empty Plan, Create Candidate Plan, Delete Plan, Edit Plan, Annotate Changeable Area, Browse Reference Material, Browse External Material. We can see most instructors use browse reference materials, edit plans, and annotate changeable areas, but there are many individual differences between participants.}
    \includegraphics[width=\textwidth]{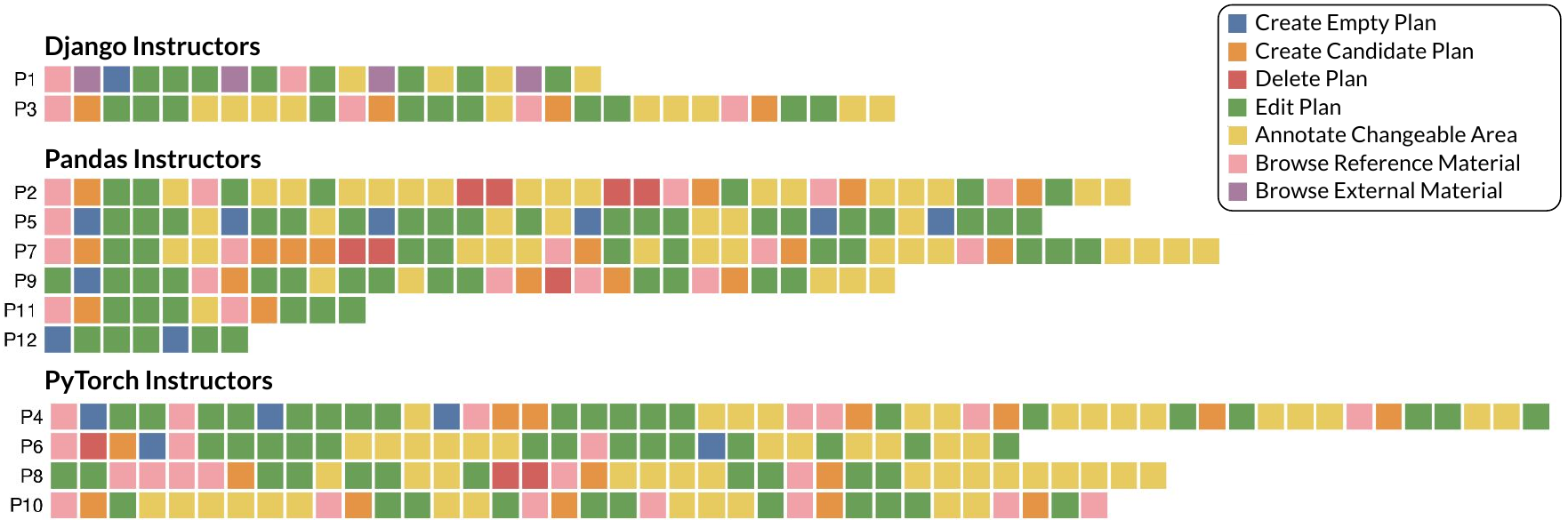}
    \caption{Trace diagram depicting participant interactions with PLAID. Each participant is shown as a horizontal line consisting of a series of actions.}
    \label{fig:trace-diagram}
\end{figure*}

\subsubsection{PLAID scaffolds instructors at multiple stages of the plan design process.}

Using our think-aloud data, post-task interviews, and trace diagrams (see \cref{fig:trace-diagram}), we noted instructors using PLAID to effectively and efficiently identify plans. 

\textbf{PLAID accelerates plan identification for instructors by providing easily navigable reference material.}
Almost all participants appreciated the example programs included as part of PLAID.
While only some participants utilized the automated suggestions based on clusters of similar code snippets (orange in \cref{fig:trace-diagram}), all participants except E12 primarily interacted with the given reference content by browsing the example programs 
and reading their short descriptions 
(pink in \cref{fig:trace-diagram}). 
Participants indicated that developing initial ideas for designing plans was the most challenging stage of plan identification. E10 said it is easier to ``\textit{derive from an existing codebase...because the sample code is the key part}'', clarifying that they believed they were more efficient when using PLAID. 
E6 appreciated the inclusion of ``\textit{readily available code snippets}'' and E11 valued the ``\textit{condensed view}'', expressing that it felt like ``\textit{going through an email inbox}''. 
Participants found it captivating to browse the list of use cases and search for key concepts. 
After they completed their timed task, E2 added ``\textit{I could keep going...I almost just want to read the list at this point.}'' 

While participants were allowed access to alternative reference content, they indicated that PLAID's technique of presenting examples was more suited to their needs. For example, multiple interviewees used ChatGPT to design plans in the baseline condition; however, they still found the process tedious. E9 communicated that the output was verbose and that it was ``\textit{quite an effort to ask even ChatGPT [for ideas]}''. E6 reported that ChatGPT split code into snippets at a different granularity than they would prefer. E5 prompted ChatGPT for ``\textit{things students struggle with when using Pandas}'' but did not find the output appropriate for beginners. ``\textit{I don't even know if I fully understand [this concept]}'', said E5.
While the queries that instructors used to prompt ChatGPT were not so different from the prompts employed as part of PLAID's pipeline,
PLAID prioritizes goal-focused information. More precisely, PLAID shows a brief natural language description of the relevant use case or candidate plan before displaying any code. Without this guidance in ChatGPT, reviewing output might be overwhelming.
Arguably, a chat interface as the only mode of interaction is challenging, as important information is very spread out and interleaved with verbose explanatory text; participants like E6 spent a long time combining code to abstract high-level ideas from multiple responses given by ChatGPT to design plans.

Participants noted weaknesses in other external reference resources as well. 
E10 found it tedious and challenging to compare inconsistent examples from multiple webpages and to find differences between these variable implementations. E8 stated, ``\textit{I know the material for this on the Internet isn't especially good}'' before they transitioned on to reference the code that they authored in the past. Even with their own code, we observed that participants needed to substantially modify their programs to meet the needs of their students. 
For example, E8 copied a snippet from code they wrote for a project and edited it, saying that ``\textit{This isn't necessarily optimal, but it's simple. That should be good for teaching material}''. E12 explained that they added structures they would otherwise not use in a complete program to help students understand (\textit{I'll do it one time, but I won't do it repeatedly}'', said E12). In contrast to other external resources, instructors referenced the documentation, often for reviewing the syntax they wanted to use in their plans.

\textbf{PLAID helps instructors create learner-friendly material by providing a structured template.}
Participants valued having a structure for designing plans. E10 found that stating explicit goals was useful for students to ``\textit{get more motivated that [they] know the purpose of learning}'' about the code. 
According to most instructors, the most advantageous part of the plan template were the changeable areas. Instructors perceived these annotations as a strategy of providing support to students. For example, E9 stated: \begin{quote}
``If I'm creating exercises in it, I'm specifying [to students] very clearly that `This is the overall intuition of the coding flow, and these are the areas that you can play with.' It kind of helps me direct the attention of the student towards the exact problem that we should be thinking about.''
\end{quote}
Most instructors used the annotation tool in PLAID to mark changeable areas in their solutions (yellow in \cref{fig:trace-diagram}). We also observed that participants who started with PLAID looked for a similar annotation mechanism in the baseline condition, pointing that there is no ``\textit{intuitive}'' (E9) way to achieve it.  

\textbf{PLAID supports the diverse iterative workflows of instructors.} We observed that instructors preferred to build high-level narratives with their plans, such as designing multiple plans that are all part of a complete program. PLAID's canvas, which shows all the in-progress plans at once, supports this behavior. E10 explained how this view was more helpful compared to the baseline: \begin{quote}
    ``The visual aspect of it [viewing boxes with only plan names], as opposed to seeing the whole thing [written-out plans in the baseline document], made it more modular, I like that abstraction. So I could focus on higher level takeaway of what I want the class to be about, instead of getting fixated of details of each [program].''
\end{quote}
Some participants imitated this process in the baseline by creating a list of initial ideas and then elaborating on each idea with other details. However, we observed that PLAID encouraged participants to keep refining and iterating at various granularities. For example, E8 designed one plan, started exploring the reference material for another, then found a concept that fit the previous plan better, and quickly went back to the previous plan to modify it as well. Similarly, E4 copied a code snippet from the reference material and made some changes, including adding a name and a goal. Then, instead of going back to the reference material or creating an empty plan, they copied the same plan and created another variation on it with small modifications.

\textbf{PLAID offers promise in introducing plan-based pedagogy to application-specific courses.}
Even though instructors did not have prior exposure to plan-focused instruction, instructors had overwhelmingly positive responses when asked about incorporating plan-based pedagogy in their instruction and using PLAID for designing plans. E9 described plan-based instruction as a ``\textit{step-by-step walkthrough of fundamental concepts}''. E11 indicated that learning about programming plans would help students retain common and important tasks that ``\textit{you can never remember the code for}''. Without any prompting, E2 and E3 even requested access to PLAID to design their upcoming courses.
However, a few participants expressed concerns about using plan-based pedagogies for instruction. For E8, plans were useful for teaching ``\textit{many small individual things}'', but they were uncertain about their usefulness when it came to combining these smaller tasks into larger projects. E1 and E5 found programming plans valuable for conceptual understanding but were hesitant to design their existing course around these structures from scratch. E12 stated that plans could be useful for some learners, but also explained that they would prefer to include executable, full programs in lecture instead.













\section{Discussion}


%
Computing education research is a leading subject area for applications of large language models (LLMs) in education, as LLMs capable of generating code at scale were made publicly available much earlier than general-purpose models like ChatGPT. 
There have been many studies that designed student-facing tools around this technology~\cite{ferdowsiValidatingAIGeneratedCode2024,jinTeachAIHow2024,kazemitabaarCodeAidEvaluatingClassroom2024,logachevaEvaluatingContextuallyPersonalized2024,yangDebuggingAITutor2024,yanIvieLightweightAnchored2024}. 
However, a main limitation explored in these works is the untrustworthy nature of LLMs, which could generate hallucinated, incorrect responses or content that is not appropriate for learners.
Arguably, instructors are particularly well-positioned to incorporate LLMs in educational workflows, as they have sufficient content knowledge to detect hallucinations and appropriate pedagogical knowledge to recognize when LLM-generated content is too technical or verbose.
However, tools that use LLMs for supporting instructors have been less common, even though there have been some successful examples (e.g.,~\cite{choiVIVIDHumanAICollaborative2024a}).


Our evaluation of PLAID confirms that LLMs can be used in the design of tools that support instructors, specifically by automating the tedious parts of instructors' workflows without undermining opportunities for them to apply their domain-specific and pedagogical expertise. 
One ubiquitous observation from our user study that validates this argument is the 
highly positive response to the LLM-generated reference material. 
All instructors valued having access to diverse and authentic examples because this accelerated an initial content collection process that would otherwise have been performed manually. 
By delegating this process to a large language model, instructors are able to focus on refining and iterating on the process of designing plans. 

Notably, instructors who attempted to use ChatGPT as an external resource in the baseline condition did not benefit as much as instructors using PLAID did, even though ChatGPT was used to generate the reference content for PLAID. 
Moreover, they found the interactions with ChatGPT to be tedious and the outputs to be verbose, even though their prompts were not so different from our queries to generate content passed into PLAID.
This highlights the importance of presenting LLM-generated content with appropriate interactions that reflect existing instructor practices. 
Thus, our findings suggest that human-in-the-loop approaches that equip instructors with preliminary content and facilitate refinement of that content are promising for the design of educational technology. 

PLAID presents encouraging results for utilizing LLM-powered tools to promote best practices and theory-informed approaches for education. Most instructors with no experience in plan-focused pedagogies were interested in using programming plans for instruction after a relatively short exposure to the concept. By streamlining the opaque and tedious process of designing a programming plan, PLAID successfully sparked interest among these instructors for adopting plan-focused pedagogies. While PLAID presently supports four application-focused domains (Pandas, Pytorch, Django, and BeautifulSoup), our versatile pipeline and design of the interface are able to support instructors in identifying plans in any domain of interest. In a sense, `robots are here'~\cite{pratherRobotsAreHere2023} for the boring and repetitive work of gathering content and organizing it into broad categories for the first draft. This delegation of work empowers instructors to focus on building overarching narratives and refining content for learners,
instead of going through a repetitive and unclear process of searching for programs that capture common patterns. Utilizing LLMs to automate repetitive information-gathering tasks, allowing instructors to use their expertise on problems with higher impact, could be an important goal for designers working on similar tools.

While most instructors saw value in programming plans, we noted that there were some logistic concerns about adopting plan-based pedagogies. Instructors who have been teaching a 

well-structured course for a long time expressed reluctance to go through the effort of a major redesign. The most positive responses came from graduate teaching assistants or instructors in the process of designing a course from scratch, who had the chance to incorporate programming plans into their courses in the first place. 
Researchers working on educational technologies for instructors should consider this hesitancy to adopt new approaches, and identify additional design considerations relevant to such instructors.






\section{Limitations and Future Work}

While our study establishes foundational steps for plan identification, a few limitations exist. Most importantly, we evaluated our system in sessions limited to 15 minutes. These sessions may not have comprehensively captured all the ways in which instructors would interact with the interface because instructor behavior could change as they become more familiar with the system. Indeed, some participants expressed that they could not explore the features enabled in the system due to time constraints and primarily used interactions that were explicitly demonstrated before the tasks. Capturing how instructors' perceptions and behaviors change over time as they interact with the system could be valuable.

In addition, our study design required the recruitment of a population with expertise in instruction and an application-focused domain. Consequently, the sample size of our participants was comparatively small ($N=12$). This sample size may not be enough to observe statistical significance in some of our tests, including the mixed effects model and the Wilcoxon signed rank tests. However, along with our qualitative observations, we find the results to still be informative and encouraging for future research. 

\edit{The quality of generated content is an important consideration for any system that utilizes large language models. An inherent limitation of LLMs in educational technology is the potential generation of inconsistent, inaccurate, or incorrect content that might be harmful to learners.  PLAID mitigates these problems with its instructor-in-the-loop design where LLM-generated content is reviewed and refined by instructors before being presented to students. By presenting many code examples and alternative suggestions for all components of a plan, PLAID encourages instructors to carefully compare content. Our user study also showed this behavior as instructors often edited parts of plans they created from examples. Moreover, our initial exploration in Section \ref{sec:llm-pipeline} showed the generated content to be syntactically accurate in most cases, suggesting that the generated content is meaningful and useful for instructors to review. The usefulness of PLAID may still be impacted by the quality of its LLM-generated content: poorer quality of the initial content may lead to lower efficiency in instructors' ability to generate final content. Future work could explore other automatic approaches to identify inaccurate content before it is presented to instructors to reduce the effort they would need to put in to refine the content to a greater extent.}

\edit{Future work should explore the generalizability of PLAID to new programming domains. In our study, we evaluated PLAID in three distinct programming domains demonstrating its usefulness beyond introductory programming. Even in complex domains like Django and PyTorch, we did not observe any statistically significant differences between instructors' performance while identifying plans. However, newer programming languages or domains may be underrepresented in the LLM training data, affecting the quality of the generated code. Examining how instructors use PLAID in these niche domains can provide important insights into the utility of PLAID and its design considerations for larger computing education research. Moreover, designing programming plans in complex domains like app development might benefit from more interactions, such as support to organize code in multiple files.}

An open question for future work is how to design novel systems that present programming plans to students and automate additional aspects of plan-based pedagogies.
Prior research on plans has shown that explicit plan-based pedagogical instruction may motivate students and support better knowledge acquisition~\cite{Cunningham_PurposeFirstProgramming_CHI-2021}. 
While the plans instructors generated in PLAID may be appropriate for use in lectures, there are many more opportunities to use these identified plans to support instruction. Our instructors proposed many interesting applications, including \textit{plans as cheatsheets}, where each plan explains a common task that students usually struggle with; \textit{plans as question generators}, where a plan with changeable areas is treated as a multiple-choice question; and \textit{plans as example generators}, creating on-demand worked examples~\cite{Atkinson_WorkedExamples_2000}.

\section{Conclusion}
\edit{In this paper, we present PLAID, a tool that assists instructors in generating programming plans in application-focused programming domains, a crucial step towards the use of plan-based pedagogies.
Such pedagogies have shown promising outcomes for introductory programming learners, but have not been applied to application-focused programming domains such as data analysis or machine learning.
Through formative interviews (N=10 educators), we identified how creating programming plans that capture high-level patterns in these domains can be challenging and tedious, even with access to AI-generated content. 
Through design workshops (N=7 educators), we derived design goals detailing how AI-generated content should be presented to instructors to make this plan creation process more efficient.
Our findings from a mixed methods within-subjects user study (N=12 educators) show that during the plan identification process, instructors experience lower cognitive demands and overall task load with PLAID, find it more satisfying to use, and prefer it over traditional approaches.
Our work not only addresses the challenges in generating programming plans for application-focused programming domains but also contributes design considerations to guide the development of ``human-in-the-loop'' AI tools.
We find that instructors can leverage LLMs to effectively author instructional material when the LLM-generated content is presented in a format that reflects best practices and reduces distractions. }
We believe PLAID can be easily extended to support plan identification in many application-focused programming domains, potentially encouraging the adoption of plan-based pedagogies at scale. Moreover, our design goals and insights from instructor-LLM interactions can inform the design of tools that support instructors in creating content in their domain of expertise. 

\begin{acks}
We thank the members of the TRAILS Lab for their insightful feedback. We also express our gratitude to the reviewers for their thoughtful and valuable suggestions. We extend our appreciation to the CS STARS program, the Siebel School of Computing and Data Science, the Campus Honors Program, and the Office of Undergraduate Research at the University of Illinois Urbana-Champaign for funding that contributed to this work.
\end{acks}
\newline
\bibliographystyle{ACM-Reference-Format}
\bibliography{bib/bib_new
, bib/background, bib/sample-base, bib/paperpile, bib/mark, bib/cps, bib/SBF-review}

\appendix
\section{Prompts Used in System Design}
\label{sec:appendix-pipeline}


\subsection{Get 100 Use Cases in Each Domain} 
\label{sec:use_case_prompt}


\begin{quote}
    \tt ``Give me 100 use cases of $\{DOMAIN\_NAME\}$. A use case describes a task you can achieve with the given library. For example, for the math library in Python, calculating the area of a circle would be an appropriately specific use case, but doing calculations would be too general. List these use cases without any comments before or after. Basically, just give me a list of use cases and no other text. In addition, these use cases should be appropriate for instruction for novices.''
\end{quote}

\subsection{Extract Code for all Generated Use Cases}
\label{sec:code_prompt}


\begin{quote}
    \tt ``Write code to use $\{DOMAIN\_NAME\}$ to achieve the task I give to you. Return the code block without any text before or after. Basically, just give me the code block and no other text. 
    
    Write code to do the following: $\{USE\_CASE\}$.''
\end{quote}

\subsection{Annotate Subgoals in Full Programs}
\label{sec:subgoals_prompt}


\begin{quote}
    \tt ``For this piece of code, instead of putting a comment for every line, could you combine comments to add subgoals? Subgoals should describe small chunks of code that achieve a task that can be explained in natural language, rather than describing the code on a single line. Each subgoal should be put as a comment before the code starts. 
    
    Here is the code: $\{FULL\_PROGRAM\}$.''
\end{quote}

\subsection{Identify Changeable Areas in Code Snippets} 
\label{sec:ca_prompt}


\begin{quote}
    \tt ``We define a domain-specific programming plan as a piece of code common in programs from a particular application area (e.g., web parsing) that achieves a specific goal. I am providing you with a piece of code. Based on this definition, can you highlight the changeable areas?  Changeable areas are the parts of the idiom that would change when it is used in different scenarios. Could you give me the exact block of code from the line that would change. Don't give me the whole line. Just give me the part of the line that would change. For example, if just the URL changes in a line, give me just the URL. Give me each of these in a code block. List these code blocks without any comments before or after. Basically, just give me a list of code blocks of code parts that would change and no other text. 
    
    Here is the code: $\{CODE\_SNIPPET\}$''
\end{quote}

\subsection{Render Names for Clusters of Code}
\label{sec:name_prompt}


\begin{quote}
    \tt ``I am giving you a cluster with comments which are the goals of some pieces of code along with the code. Come up with a name for this cluster of plans. Programming plans are pieces of code common in programs from a particular application area that are used to achieve a given goal. A name is reflective of what that plan achieves. So produce a name that would help me understand what goal the code  will be achieving. To reiterate, be very specific, and consider what each subgoal is doing. Do not consider the context. Just consider what the code is doing. Return the result to me in the form of the following string, "Name: ". 
    
    Here is the cluster: $\{PROGRAMS\_IN\_CLUSTER\}$''
\end{quote}

\end{document}